\documentclass[sn-mathphys,Numbered]{sn-jnl}

\usepackage{graphicx}%
\usepackage{multirow}%
\usepackage{amsmath,amssymb,amsfonts}%
\usepackage{amsthm}%
\usepackage{mathrsfs}%
\usepackage[title]{appendix}%
\usepackage{xcolor}%
\usepackage{textcomp}%
\usepackage{manyfoot}%
\usepackage{booktabs}%
\usepackage{algorithm}%
\usepackage{algorithmicx}%
\usepackage{algpseudocode}%
\usepackage{listings}%
\usepackage{graphics}
\usepackage{graphicx}
\usepackage{subfig}
\usepackage{tabularx}

\theoremstyle{thmstyleone}%
\newtheorem{theorem}{Theorem}
\newtheorem{proposition}[theorem]{Proposition}%

\theoremstyle{thmstyletwo}%
\newtheorem{example}{Example}%

\theoremstyle{thmstylethree}%
\newtheorem{definition}{Definition}%

\begin{document}

\title[title]{Dissipative Soliton Resonance: Adiabatic Theory and Thermodynamics}

\author*[1]{\fnm{Vladimir L.} \sur{Kalashnikov}}\email{vladimir.kalashnikov@ntnu.no}

\author[1]{\fnm{Alexander} \sur{Rudenkov}}\email{alexander.rudenkov@ntnu.no}
\equalcont{These authors contributed equally to this work.}

\author[2,3]{\fnm{Evgeni} \sur{Sorokin}}\email{evgeni.sorokin@tuwien.ac.at}
\equalcont{These authors contributed equally to this work.}

\author[1,3]{\fnm{Irina T.} \sur{Sorokina}}\email{irina.sorokina@ntnu.no}
\equalcont{These authors contributed equally to this work.}

\affil*[1]{\orgdiv{Department of Physics}, \orgname{Norwegian University of Science and Technology}, \city{Trondheim}, \postcode{7491}, \country{Norway}}

\affil[2]{\orgdiv{Institut fuer Photonik}, \orgname{Vienna University of Technology}, \orgaddress{\street{Gusshausstrasse 27/387}, \city{Vienna}, \postcode{A-1040}, \country{Austria}}}

\affil[3]{\orgname{ATLA lasers AS}, \orgaddress{\street{Richard Birkelands vei 2B}, \city{Trondheim}, \postcode{7034}, \country{Norway}}}


\abstract{We present the adiabatic theory of dissipative solitons (DS) of complex cubic-quintic nonlinear Ginzburg-Landau equation (CQGLE). Solutions in the closed analytical form in the spectral domain have the shape of Rayleigh-Jeans distribution for olive a positive (normal) dispersion. The DS parametric space forms a two-dimensional (or three-dimensional for the complex quintic nonlinearity) master diagram connecting the DS energy and a universal parameter formed by the ratio of four real and imaginary coefficients for dissipative and non-dissipative terms in CQGLE. The concept of dissipative soliton resonance (DSR) is formulated in terms of the master diagram, and the main signatures of transition to DSR are demonstrated and experimentally verified. We show a close analogy between DS and incoherent (semicoherent) solitons with an ensemble of quasi-particles confined by a collective potential. It allows applying the thermodynamical approach to DS and deriving the conditions for the DS energy scalability.}




\maketitle

\section{Introduction}\label{sec1}

Many recent scientific breakthroughs in various fields were made possible by using ultrashort pulse lasers and understanding how a dissipative soliton (DS) forms and works. DS is a stable and localized pattern with different levels of coherence, which arises in a nonlinear system far from equilibrium due to energy loss or gain. The DS concept applies in diverse scientific areas, such as cosmology, optics, physics, biology, and medicine \cite{DS,ankiewicz2008dissipative,purwins2010dissipative}. Due to the nonequilibrium of a system, DS needs to exchange energy with the environment in a well-organized way. This energy flow shapes the internal structure of DS, which allows the energy to be redistributed within it. In this sense, a DS is a simple version of a cell. A complex internal structure of DS affects its behavior and can lead even to turbulence that links DS to a family of incoherent or semicoherent solitons \cite{picozzi1,picozzi2}. The variety of phenomena that optical DS can mimic, such as turbulence, noise, and rogue waves \cite{grelu2012dissipative,dudley2014instabilities}, makes them useful for studying nonlinear systems and thermodynamics far from equilibrium. Moreover, they offer us powerful and flexible methods for simulating, computing, and analyzing large and rare data sets that can be applied to different fields of science, technology, and medicine. Moreover, DS phenomenology allows us to use powerful and adjustable methods for metaphorical computing and modeling processes from distant fields of science \cite{editorial}.

Despite the evident fact that DS is a classical field structure due to the large $k-$mode occupation number $n_k \gg1$ and strong entanglement with an environment, the non-trivial DS composition enhanced by spectral-temporal condensation and the resonant enhancement of sensitivity to perturbations throw a bridge across microscopic and mesoscopic physics and put a question about the quantum theory of DS \cite{kaup1975exact,lai1989quantum}. The latter is especially important in view of the close analogy between coherent structures in photonics and Bose-Einstein condensate (BEC) \cite{klaers2010bose,sun2012observation,sob2013bose,bagnato2015bose,kalashnikov2021metaphorical,bloch2022non}. A scalable formation of coherent condensate phase in the DS form was named \textit{dissipative soliton resonance} (DSR) \cite{chang2008dissipative}.

The theoretical workhorse in the above-mentioned endeavors was the \textit{complex nonlinear Ginzburg-Landau equation} \cite{van1992fronts,haus1992analytic,aranson2002world,Ginzburg2009,ferreira2022dissipative}, which is akin the \textit{nonlinear Schr\"{o}dinger equation} \cite{NSE,Liu2019} so that some terminological confusing could appear. It is possible to divide these terms conditionally based on the object under consideration: DS solutions of the complex nonlinear Ginzburg-Landau equation considered in this work have no non-dissipative limit, i.e., they exist in the non-soliton sector of the nonlinear Schr\"{o}dinger equation, where $\gamma/\beta <0$, with $\gamma$ and $\beta$ being the coefficients of the imaginary terms characterizing nonlinearity and kinetic (dispersion, diffraction, or kinetic energy) parameters in a system, respectively. The same concerns the \textit{Gross-Pitaevskii equation}, which is actively exploited in the studies on BEC \footnote{There is inexhaustible literature on this topic. Therefore, we limit ourselves to a single review citation \cite{carretero2008nonlinear}}. In this work, we consider a region of blue normal group-delay dispersion (GDD in an optical context). The nonlinear gain and spectral filtering \cite{bale2008spectral} (or viscous friction \cite{leblond2016dissipative}) are absolutely necessary for the existence of this type of DS blue emerging in a dissipative non-conservative system.

The article is organized in the following way. Firstly, we expose the adiabatic theory of DS of the complex cubic-quintic nonlinear Ginzburg-Landau equation (CQGLE) and shortly characterize their properties with focus on the DS spectra, which have a shape of the Rayleigh-Jeans distribution in the simplest case. Then, the DS parametric space is formulated in terms of the master diagram, which is two-dimensional for the reduced CQGLE and three-dimensional for the complete CQGLE. The concept of \textit{dissipative soliton resonance} (DSR) \cite{chang2008dissipative} is formulated using the adiabatic theory. Finally, we consider, in short, the thermodynamics of the strongly chirped DS using an ideology of the in-(semi-)coherent soliton theory \cite{picozzi1}.

\section{Adiabatic theory of dissipative solitons}\label{sec2}

Let us consider the (1+1)-dimensional CQGLE, which describes an evolution of the field envelope $a(z,t)$ in the following form \cite{grelu2012dissipative,ferreira2022dissipative}:
\begin{align}
\frac{\partial}{\partial z} a\! \left(z,t\right)=-\Sigma a\! \left(z,t\right)+\left(\alpha+\mathrm{i} \beta\right) \frac{\partial^{2}}{\partial t^{2}} a\! \left(z,t\right)+ \nonumber \\
+ \left(\kappa-\mathrm{i} \gamma\right)P a\! \left(z,t\right)-\left(\kappa \zeta+\mathrm{i} \chi\right) P^2  a\! \left(z,t\right).\label{eq:CNGLE}
\end{align}

\noindent Here, we consider $z$ as an evolution coordinate (e.g., propagation distance in a laser/waveguide or time in BEC), and $t$ is the local (co-moving) time coordinate (or spatial coordinate in a planar waveguide or BEC). $P=|a\left(z,t\right)|^2$ in Eq. (\ref{eq:CNGLE}). The $\beta$-term describes an action of GDD. The anomalous GDD $\beta <$0 corresponds to the diffraction term for a planar waveguide or the kinetic energy of bosons. Table \ref{tab:table1} illustrates an example of physical correspondence between photonics and BEC physical interpretation of the terms in Eq. (\ref{eq:CNGLE})\footnote{We do not consider the effect of spatial confinement but it could be easily grasped by Eq. (\ref{eq:CNGLE}).}.

\begin{table}
 \caption{Correspondences between photonics and BEC \cite{kalashnikov2021metaphorical}}
  \centering
  \begin{tabularx}\columnwidth{X p{1ex} X}
   \toprule
    Laser    & & BEC \\
    \midrule
    Propagation coordinate $z$ &$\leftrightarrow$& time $T$\\
    Pulse local time $t$ &$\leftrightarrow$& third spatial coordinate $z$\\
    Diffraction + anomalous GDD ($\beta$)     &$\leftrightarrow$& boson kinetic energy\\
    Kerr-nonlinearity ($\gamma$)  &$\leftrightarrow$& boson attractive colliding potential\\
        Non-selective loss ($\Sigma$) &$\leftrightarrow$& homogeneous BEC  dissipation\\
        Spectral dissipation ($\alpha$) &$\Leftrightarrow$& ``kinetic cooling''\\
        Saturable nonlinear gain ($\kappa$, $\zeta$) &$\Leftrightarrow$& condensation from finite basin
  \end{tabularx}
  \label{tab:table1}
\end{table}

Below, we will consider the case of $\beta >$0 (\emph{normal GDD}) that breaks the above analogy between temporal phenomena in optics and spatial phenomena in the waveguide and condensed-matter physics.

The nonlinear non-dissipative terms $\gamma>$0 and $\chi$ describe the self-phase modulation (SPM) (self-focusing or attracting boson interaction in the spatial domain), which is saturable ($\chi<$0) or growable ($\chi>$0) with power. In a laser, the quintic nonlinear term $\chi$ can appear, for instance, due to the mode size variation caused by the self-focusing.

The dissipative terms in Eq. (\ref{eq:CNGLE}) describe $\sigma$ -- a saturated net-loss defined by interaction with a finite basin causing loss and gain, which is saturated by the full field energy $\int \left| a \right|^2 dt$. $\alpha$ -- a spectral dissipation (``kinetic cooling'' \cite{kalashnikov2021metaphorical}). In a laser, this parameter equals the squared inverse bandwidth of a spectral filter, which is formed by the finite gain bandwidth of the active medium, spectral filters, mirror coatings, etc. $\kappa$ and $\zeta$ -- the saturable nonlinear gain (self-amplitude modulation, SAM) providing excessive but top-bounded ($\zeta>0$) gain for the higher peak power signal over noise.     

The well-known exact soliton-like solution of (\ref{eq:CNGLE})

\begin{equation}\label{eq:exact}
    a(z,t) = \sqrt{\frac{\mathfrak{A}}{\cosh(t/T)+\mathfrak{B}}}\exp[-i\psi/2 \ln(\cosh(t/T))-iqz],
\end{equation}
\noindent was widely explored \cite{moores1993ginzburg,soto1997pulse,renninger2008dissipative}\footnote{The solution of a dissipation-free version of (\ref{eq:CNGLE}) was presented in \cite{pushkarov1979self}.}.
Here $\left\{ \mathfrak{A},\ \mathfrak{B}, \ T, \ q \right\} \in \Re$. Innovative methods to derive the solutions of (\ref{eq:CNGLE}) based on the data-driven approach are presented in \cite{yin2023dynamic}.

The new insights into the CQGLE world could be provided by the approximated methods based on the perturbative method \cite{malomed1990kinks,malomed1987evolution}, Lagrangian approach, and method of moments \cite{ankiewicz2007dissipative,ankiewicz2008comparison}, etc. Here, we will build on the \emph{adiabatic theory of the strongly chirped DS}\footnote{This theory was first developed in \cite{podivilov2005heavily}, and its further applications can be found in \cite{kalashnikov2009chirped,kalashnikov2009chirped2,kharenko2011highly,skiadas2017handbook}. A similar approach was suggested in \cite{ablowitz2009solitons}.} based on the following propositions:

\begin{proposition}
The nondissipative terms dominate strongly over the dissipative ones in Eq. (\ref{eq:CNGLE}): $\alpha/\beta \ll 1 \wedge \kappa/\gamma \ll 1$. 
\end{proposition} 
One must note that the first two conditions do not require the quasi-homogeneous approximation $L_{nl} \ll L_{l}$, where $L_{nl}\propto 1/\left\{ \gamma, \kappa \right\}$ and $L_{l}\propto 1/\left\{ \alpha, \beta \right\}$ are the effective nonlinear, and linear lengths in (\ref{eq:CNGLE}), respectively \cite{picozzi1}. However, as it will be shown below, the \emph{large DS chirp} $\psi$ (i.e., DS phase inhomogeneity) could play a role of the ``paraxial approximation'' \cite{picozzi2} connecting the characteristic ``correlation lengths'' in the time ($T$, DS width) and spectral ($\Delta$, DS spectral width) domains: $T \Delta \simeq \psi \gg 1$. One has to note that the perturbative analysis of the soliton-like solutions of CQGLE under the conditions of this Proposition was considered in \cite{malomed1990kinks}. 

\begin{proposition}
    $C \equiv  \frac{\alpha}{\beta} \times \frac{\gamma}{\kappa} \simeq 1.$ 
\end{proposition}
This Proposition means proximity to the \emph{soliton} or \emph{potential condition} \cite{fischer1} (although the sign before $\beta$ in (\ref{eq:CNGLE}) is inverse relatively that for a familiar nonlinear Schr\"odinger equation!). This could allow conjecturing that the steady-state probability distribution for a partially coherent DS is Gibbs-like. We note that the last conjecture is not a pre-assumption for further analysis, but, as it will be shown below, it means proximity to the DSR condition \footnote{The definition of DSR will be given below.}.         

\begin{proposition}
    \textbf{Adiabatic approximation}: field envelope $\sqrt{P\! \left(t\right)}$ evolves with $t$ slowly in comparison with the instant phase  $\varphi\left(t\right)$ change. 
\end{proposition}
In this Proposition, we assume the standard traveling wave ansatz\footnote{Such ansatz corresponds to the slowly varying envelope approximation when the DS width is much larger than the wave period.}:
\begin{equation}
a\! \left(z,t\right)=\sqrt{P\! \left(t\right)}\, {\mathrm e}^{\mathrm{i} \varphi\left(t\right)-\mathrm{i} q z},\label{eq:ansatz}
\end{equation}
\noindent where $P\! \left(t\right)$ is a slowly-varying DS power, $\varphi\left(t\right)$ is an instant phase, and $q$ is a wavenumber (propagation constant). Formally, this means that DS is ``long'' in comparison with the characteristic scale $\sqrt{\beta}$ so that one may  omit the terms $\propto \frac{d^2 \sqrt{P}}{dt^2}$ after substitution of (\ref{eq:ansatz}) into (\ref{eq:CNGLE}). After such a substitution and using the first and third propositions, one has\footnote{We follow the calculations in \cite{kalashnikov2009chirped}, and the corresponding algebra can be found in \cite{maple1}.}:

\begin{align}
    \beta \Omega(t)^2 = q - P(t) (\gamma + P(t) \chi), \label{eq:f1}\\
    P\left(t\right) \kappa -P\left(t\right)^{2} \kappa \zeta=\sigma+\alpha\Omega\left(t\right)^{2}+\frac{\beta  \Omega \left(t\right) \frac{d}{d t} P\left(t\right)}{P\! \left(t\right)}+\beta\frac{d}{d t} \Omega \left(t\right), \label{eq:f2}
\end{align}
\noindent where $\Omega(t) = d \varphi(t)/dt$ is an instant frequency deviation.

\subsection{DS having a $\chi \to 0$ limit}

Eq. (\ref{eq:f1}) allows us to obtain the expressions for the DS envelope $P(t)$:

\begin{align}
    P\! \left(t\right)=\frac{1}{2} \frac{-\gamma+\sqrt{\gamma^{2}+4 \chi( q-\beta \Omega\! \left(t\right)^{2})}}{\chi},\label{eq:P1} \\
    P\! \left(t\right)=-\frac{1}{2} \frac{\gamma+\sqrt{\gamma^{2}+4 \chi (q-\beta \Omega\! \left(t\right)^{2})}}{\chi}. \label{eq:P2}
\end{align}

Eq. (\ref{eq:P1}) has the limit of $\chi \to 0$ corresponding to DS of the reduced CQGLE: $\gamma P\! \left(t\right)=q-\beta \Omega\! \left(t\right)^{2}$ \cite{podivilov2005heavily}, and we will concentrate on the solution (\ref{eq:P1}) below.

The temporal localization of DS, i.e., $\lim_{t\to \pm \infty }P(t)=0$, gives the expression for the maximal instant frequency, i.e., the \emph{cut-off frequency} $\Delta$:

\begin{equation} \label{eq:cutoff}
    \Delta^2 = q/\beta.
\end{equation}
\noindent This expression and Proposition 2 expose that the DS considered by us belongs to the normal GDD range $\beta>0$.

Eqs. (\ref{eq:P1},\ref{eq:cutoff}) allow  excluding $P(t)$ from (\ref{eq:f2}) that leads to the expression for the instant frequency deviation:

\begin{align} \label{eq:tchirp}
    \frac{d}{d t} \Omega \left(t\right)=-\frac{\left(\sigma +\alpha  \Omega \! \left(t\right)^{2}+\frac{\kappa}{4\chi^{2}} \left(\gamma -A\right) \left(2 \chi+\zeta\cdot \left(\gamma-A\right)\right)\right) \left(\gamma -A\right) A}{\beta  \left(4 \chi  \beta  \Omega \! \left(t\right)^{2}+\left(\gamma -A\right) A\right)}, \nonumber \\
A=\sqrt{\gamma^{2}+4\beta\chi\left(\Delta^{2}-\Omega \left(t\right)^{2}\right)}\ \ \& \ \ \Omega(t)^2 \le \Delta^2.
\end{align}

Then, the cut-off frequency $\Delta$ can be obtained after some algebra from Eqs. (\ref{eq:f1},\ref{eq:f2},\ref{eq:P1},\ref{eq:tchirp}):

\begin{equation} \label{eq:cutoff2}
    \frac{\zeta \beta}{\gamma}\Delta^{2}=\frac{\left(\frac{2 \left(3+\frac{\left(C+4\right)}{b}\right)\left(2+\frac{\left(C+3b\right)}{2}\pm \sqrt{\left(2-C\right)^{2}-16 \Sigma\left(\frac{C}{b}+1\right)}\right)}{\frac{C}{b}+1}-3\left(C+3b\right)-\frac{32 \Sigma}{b}-12\right)}{16\left(\frac{C}{b}+1\right)},
\end{equation}
\noindent where the new combined parameters are introduced: \emph{control parameter} $C = \alpha \gamma/\beta \kappa$ (see Proposition 2), \emph{relative quintic parameter} $b=\gamma \zeta/\chi$, and \emph{composite net-loss parameter} $\Sigma=\sigma \zeta/\gamma$. 

\begin{figure}[h]%
\centering
\includegraphics[width=0.8\textwidth]{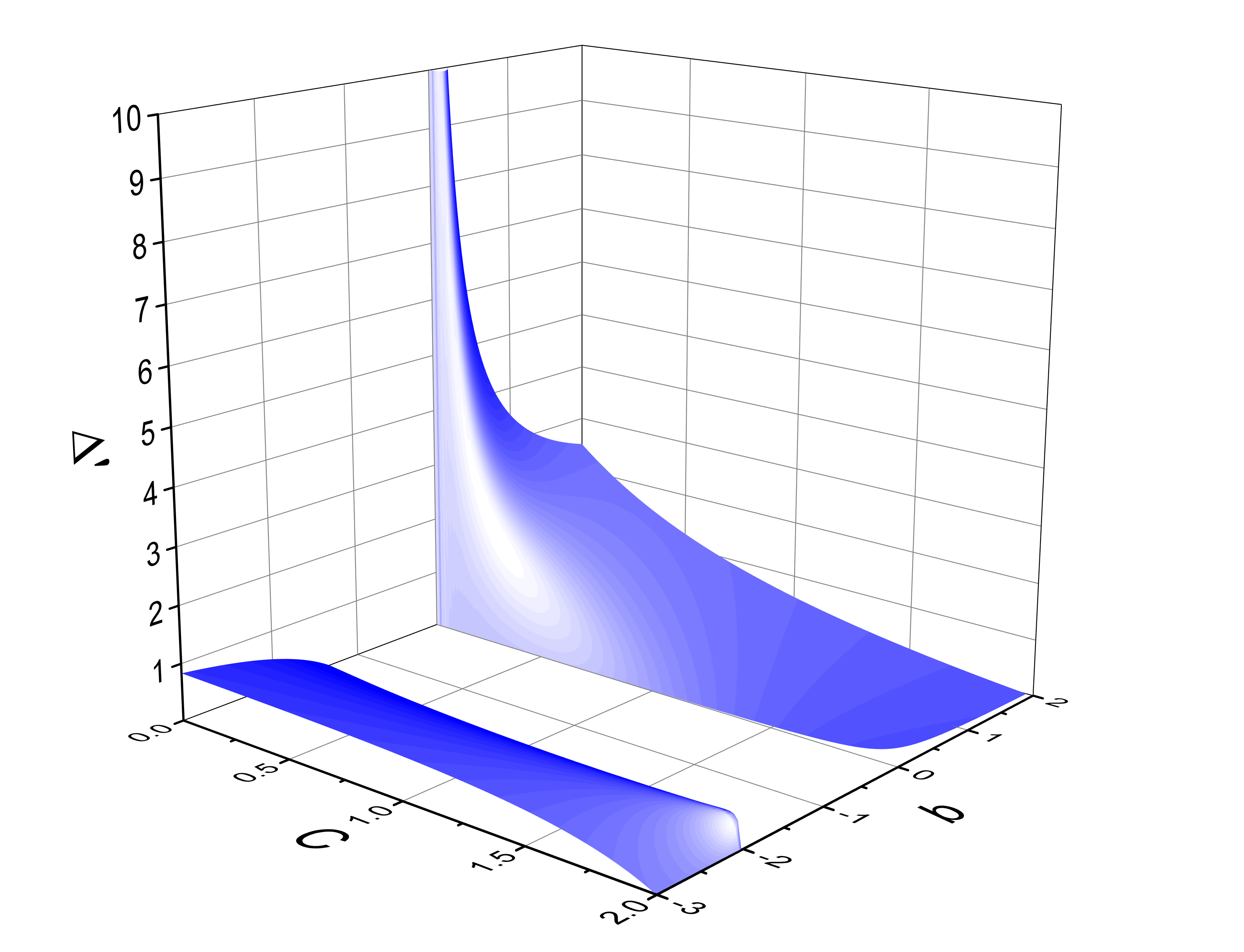}
\caption{The dimensionless cut-off frequency $\Delta'$ in dependence on the ($C,\ b$)-parameters for the $(+)$-branch of DS on the stability threshold $\Sigma=0$.}
\label{fig:fig1}
\end{figure}

The $\pm$ signs in Eq. (\ref{eq:cutoff2}) denote two branches of DS solutions. The crucial characteristic of these branches is their stability against a vacuum excitation, which means $\sigma \ge  0$ in Eq. (\ref{eq:CNGLE}). For the $(+)$-branch, the squared dimensionless cut-off frequency $\Delta'^2=\zeta \beta \Delta^2/\gamma$ on the stability threshold $\sigma=0$

\begin{equation}\label{cutoffP}
   \Delta'^2 = \frac{1}{4} \frac{b C \left(2-c\right) \left(C+3 b+4\right)}{\left(C+b\right)^{2}}
\end{equation}

\noindent is shown in Fig. \ref{fig:fig1}, where the existence range is defined as $\mathbf{C\in ]0,2]}\ \&\ \mathbf{b>0}\mathbf{\bigcup} \mathbf{b<-C/3-4/3}$. The $(-)$-branch is detached from the unstable vacuum within these regions in the sense that $\sigma>0$ for it.

\begin{figure}[h]%
\centering
\includegraphics[width=0.8\textwidth]{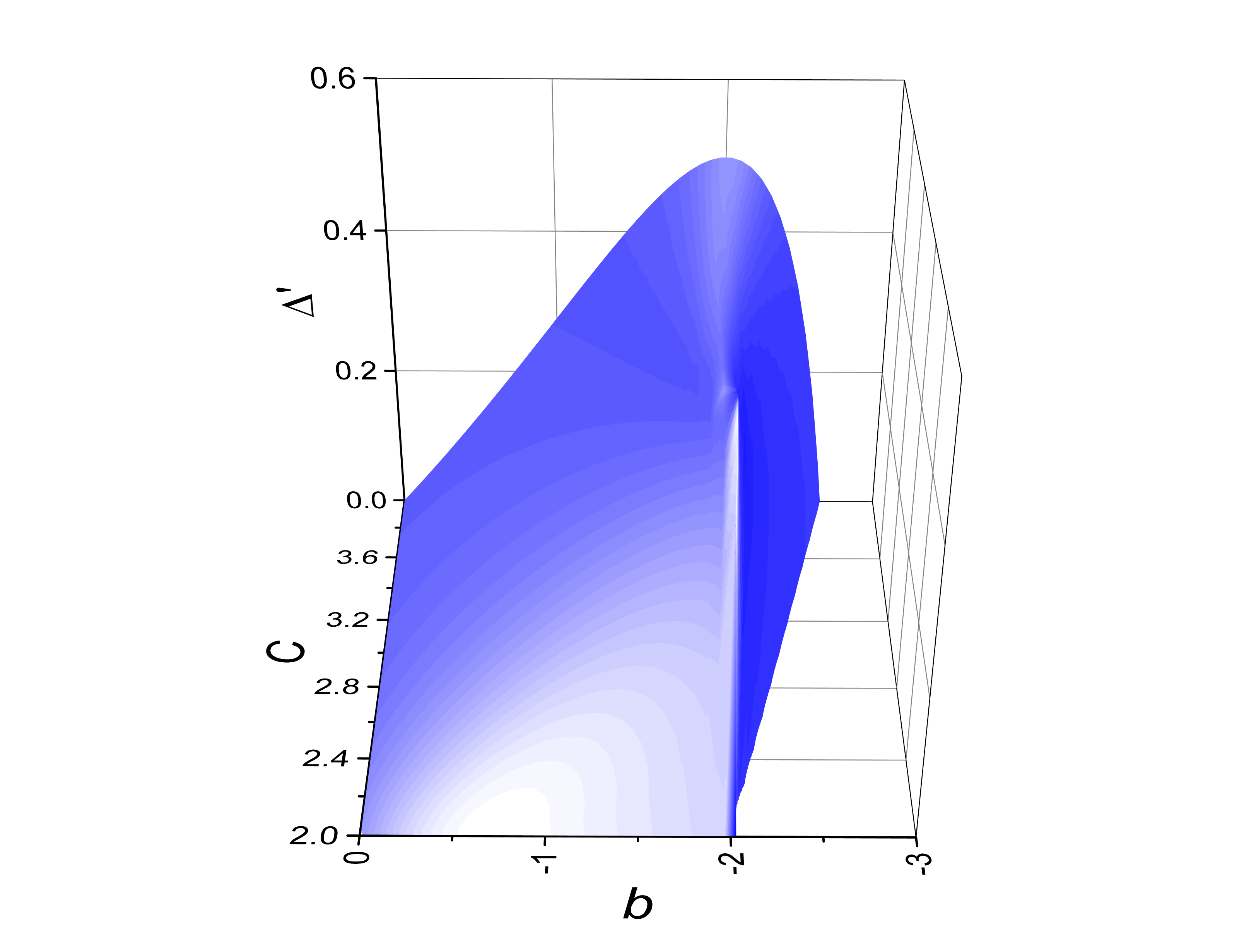}
\caption{The dimensionless cut-off frequency $\Delta'$ in dependence on the ($C,\ b$)-parameters for the $(-)$-branch of DS on the stability threshold $\Sigma=0$.}
\label{fig:fig2}
\end{figure}

The $(-)$-branch of the solution (\ref{eq:cutoff2}) could be ``connected'' with the region of unstable vacuum in the sense that it has a marginally stable limit $\sigma \to 0$ (Fig. \ref{fig:fig2}). In this case, its existence domain corresponds to the ``enhancing'' self-phase modulation, i.e., $b<0$:

\begin{equation}
    \Delta'^2 = -\frac{1}{4} \frac{b C \left(c-2\right) \left(C+3 b+4\right) }{\left(C+b\right)^{2}}
\end{equation}

\noindent that lies out of Proposition 2, and there exists no for $b \to \infty$, i.e., it is not connected to the reduced cubic-quintic DS with $\chi \to 0$. Therefore we will not consider it so that $b>0$ below except for Fig. (\ref{fig:fig9}).

\begin{figure}[h]
\centering
\includegraphics[width=0.8\textwidth]{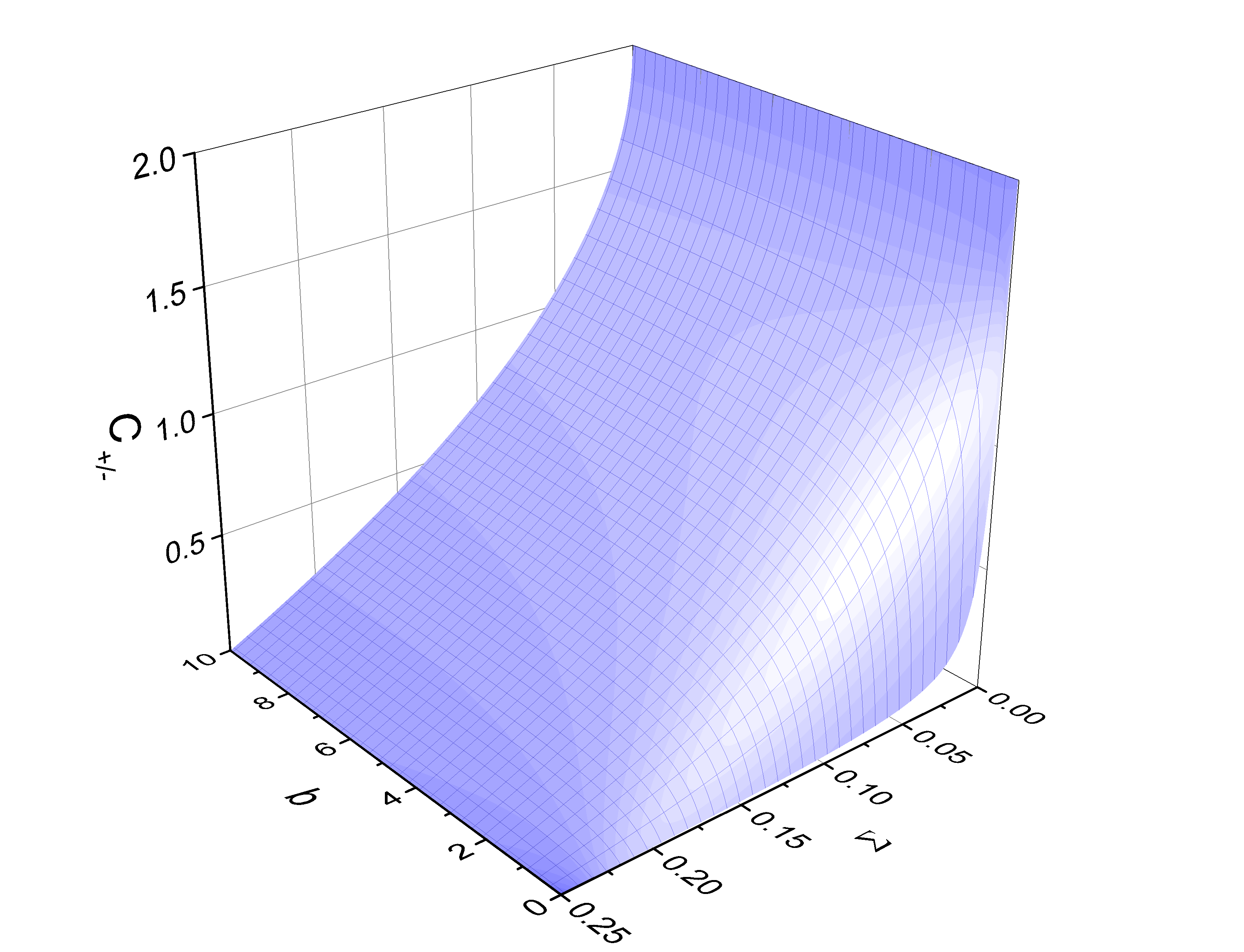}
\caption{The $\Sigma_{\pm}$-parameter dividing the $(+)$ and $(-)$ branches of DS in dependence on ($\Sigma,\ b$). Only the case of $b>0$ is illustrated.}
\label{fig:fig3}
\end{figure}

The DS branches are divided by a surface (Fig. \ref{fig:fig3} for a positive $b$):

\begin{equation} \label{eq:div}
    C_{\pm } = \frac{8 \Sigma +2 b-4 \sqrt{\Sigma  \left(b^{2}+4 \Sigma +2 b\right)}}{b},
\end{equation}

\noindent so that, for $(+)$-branch, the net-loss parameter is confined within the regions of $0<\Sigma<\Sigma_{\pm }$ if $b>0$\ \&\ $\Sigma>\Sigma_{\pm }$ if $b<-2$. Also, let's note that $\lim_{c \to 0} \Sigma=0.25$. 

\begin{figure}
  \centering
  \begin{tabular}{@{}c@{}}
    \includegraphics[scale=0.3]{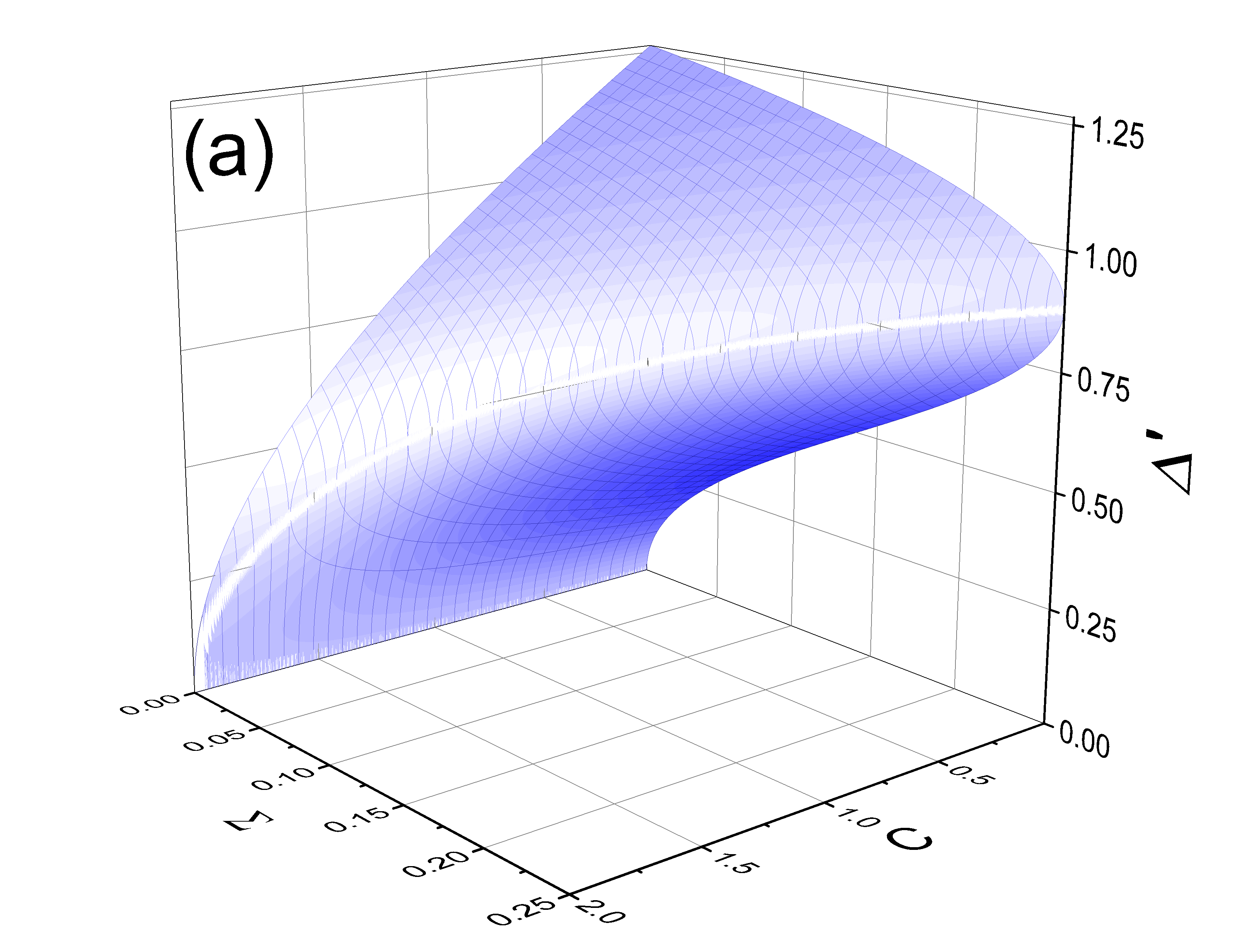} \\[\abovecaptionskip]
  \end{tabular}


  \begin{tabular}{@{}c@{}}
    \includegraphics[scale=0.3]{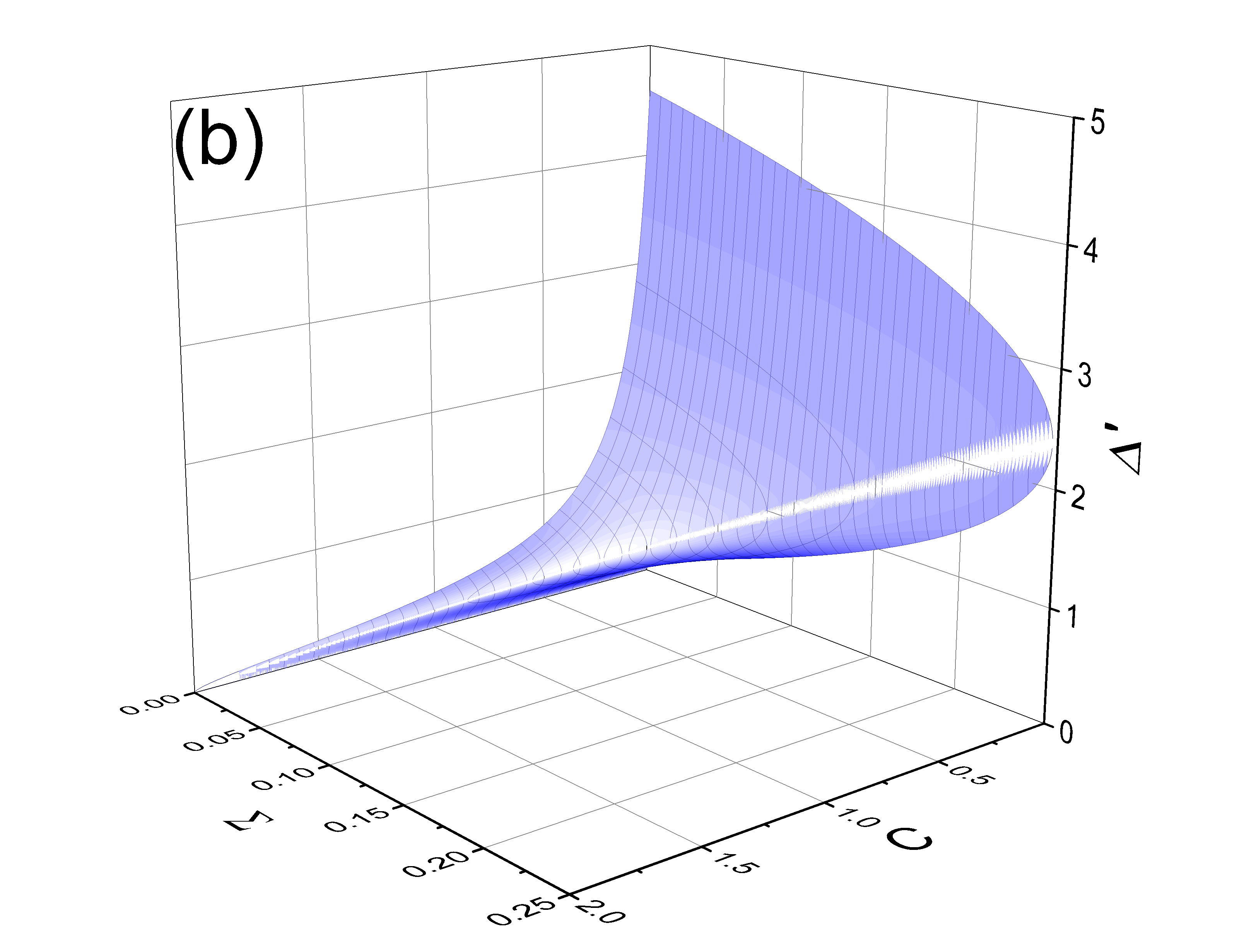} \\[\abovecaptionskip]
  \end{tabular}

  \caption{Dimensionless cut-off frequency $\Delta'$ in dependence on the net-loss $\Sigma$ and the control parameter $C$ for both branches of the DS solution: upper/bottom sheets correspond to the ($+$)/($-$)-branches, respectively. $b=20$ (a) and 0.1 (b).}\label{fig:fig4}
\end{figure}

The dependencies of $\Delta'$ on the control parameter $0\le C \le 2$ and the net-loss $\Sigma$ for both branches of DS are shown in Fig. \ref{fig:fig4}. These branches coincide on the surface shown in Fig. \ref{fig:fig3}.  The $(+)$-branch has a more significant cut-off frequency that, as it will be shown later, corresponds to the DS ``fine-graining'', its chirp growth, and a minimization of the pulse width $T_c$ after its compression by a de-chirping would be $T_c \propto 1/\Delta$. 

\begin{figure}
    \centering
    \includegraphics[scale=0.3]{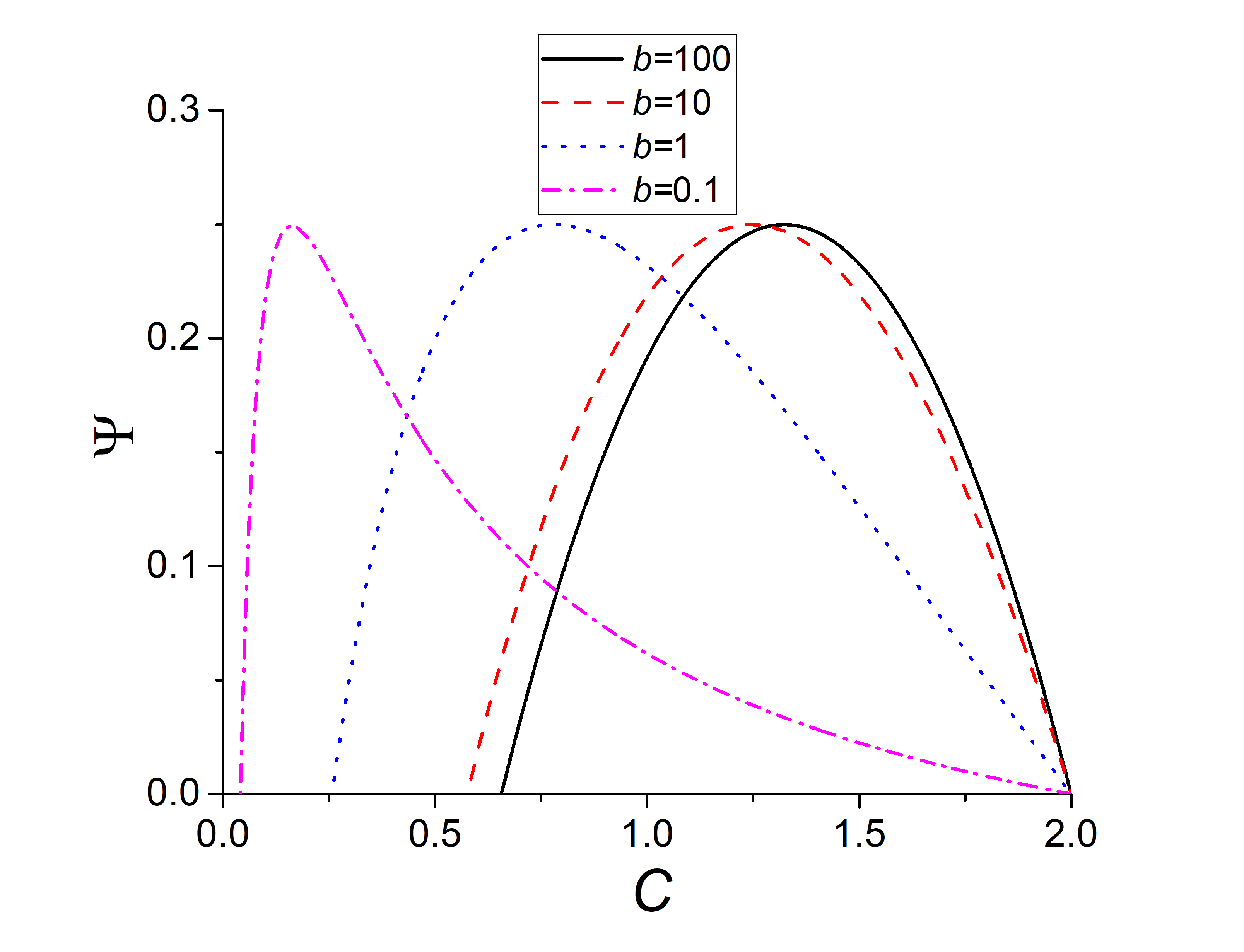}
    \caption{The dimensionless chirp $\Psi(t=0)$ of $(+)$-branch in dependence on ($C,\ b$)-parameters on the stability threshold $\Sigma=0$.}
    \label{fig:fig5}
\end{figure}

As it was noted above, the important parameter characterizing DS is the chirp which we define as $\Psi = (\beta \gamma/\kappa)\times (d \Omega'(t)/dt')$ (we use the normalization for time and frequency as above):

\begin{equation}\label{fig:chirp}
    \Psi=-(\Sigma+\frac{b^2}{4}(1-\sqrt{1+\frac{4\Delta'^2}{b}})(1+\frac{2}{b}-\sqrt{1+\frac{4\Delta'^2}{b}})),
\end{equation}

\noindent where a zero frequency deviation at $t=0$, i.e., $d\varphi(t)/dt|_{t=0}=0$, is taken into account in (\ref{eq:tchirp}). The chirps in the DS center ($t=0$) on the stability threshold $\Sigma=0$ ($"+"$ branch) are shown in Fig. \ref{fig:fig5} in dependence on $C$-parameter and positive $b$. Interestingly, the central chirp ($t=0$) tends to zero for some minimal $C$ (e.g., for $C=2/3$ and $\chi \to 0$).

We assume the large chirp in accordance with Proposition 1. That means a fast variation of the DS phase $\phi (t)$ with the time that allows applying the stationary phase approximation \cite{bleistein1975asymptotic,podivilov2005heavily,kalashnikov2009chirped}. One may assume that the Fourier transform of Eq. (\ref{eq:ansatz})

\begin{equation} \label{eq:fourier}
    e(\omega')=1/\sqrt{2}\int_{-\infty }^{\infty }\sqrt{b(\sqrt{1+4(\Delta'^2-\Omega(t)'^2)/b \ C}-1)}\exp[i(\varphi(t)-\omega' t)]dt
\end{equation}

\noindent is dominated by the contribution from the stationary points $a$ where $d\varphi(t)/dt|_{t=a}=0$ so that the leading term in the Taylor expansion of the phase is $(d^2\varphi(t)/dt^2|_{t=a})t^2/2$. Thus, Eqs. (\ref{eq:fourier}, \ref{eq:tchirp}) after some algebra lead to the expression for a complex spectral amplitude:

\begin{equation} \label{eq:sprofile}
    e(\omega')=\frac{\sqrt{\pi\, b \left(B-1\right)}\, {\mathrm e}^{\frac{\frac{\mathrm{I}}{2} \left((B -1) b +4(2 \omega'^{2}-\Delta'^{2})\right) \omega'^{2}}{B b \left((B-1)(\Sigma   +C \omega'^{2}+b  +b^{2}) -3 b (\Delta'^{2}- \omega'^{2})-2 (\Delta'^{2}- \omega'^{2})+b B (\Delta'^{2}-\omega'^{2}\right)}} }{\sqrt{\frac{\mathrm{I} B b \left((B-1)(\Sigma   +C \omega'^{2}+b  +\,b^{2}) -3 b (\Delta'^{2}- \omega'^{2})-2 (\Delta'^{2}- \omega'^{2})+b B (\Delta'^{2}-\omega'^{2}\right)}{(B -1) C b +4(2 \omega'^{2}-\Delta'^{2})}}} \mathcal{H}\left(\Delta'^2-\omega'^2\right),
\end{equation}

\noindent where $B=\sqrt{\frac{4 \Delta'^{2}+b-4 \omega'^{2}}{b}}$ and $\mathcal{H}$ is a Heaviside function. From Eq. (\ref{eq:sprofile}), one may obtain the DS spectral profile:

\begin{equation} \label{eq:spectrum}
    s\! \left(\omega'\right) \equiv \left| e(\omega') \right|^2=\frac{\left(A-1\right) \pi \left(\left(A-1\right) b+4 (2\omega'^{2}- \Delta'^{2})\right) \mathcal{H}\left(\Delta'^2-\omega'^2\right)}{A \left(\left((\Sigma+C \,\omega'^{2}+ b+b^{2})+b (\Delta'^{2}- \omega'^{2})\right) \left(A-1\right)-2 \left(\Delta'^2-\omega'^2\right) \left(b+1\right)\right)}.
\end{equation}

\begin{example}
Eq. (\ref{eq:spectrum}) has a limit $b \to \infty$ (i.e., $\chi \to 0$), which is important for further consideration. In the dimensionless form and after factorization (see Appendix), it looks as \cite{podivilov2005heavily}:
\begin{equation} \label{eq:RJ}
    s(\omega')=\frac{6 \pi\mathcal{H}\left(\Delta'^2-\omega'^2\right)}{\Xi'^2+\omega'^2},
\end{equation}
\noindent This spectrum has the form of the Rayleigh-Jeans distribution with a negative chemical potential:
\begin{equation} \label{eq:chemical}
    \Xi'^{2}=-\frac{5}{3} \Delta'^{2}+C+1.
\end{equation}
Such a similarity is not only formal and has substantial consequences (see below, and Refs. \cite{picozzi1,picozzi2,wu2019thermodynamic}).
\end{example}

\subsection{DS temporal profiles and spectra}

Unlike the exact solution (\ref{eq:exact}) with the fixed parameters\footnote{fixed in the sense of \cite{soto1997pulse,renninger2008dissipative}, when the restriction on the four free parameters of Eq. (\ref{eq:CNGLE}) (i.e., $\tau$, $\kappa$, $\zeta$, and $\chi$ in our notations) are imposed.}, the adiabatic approximation provides an approximated solution but without the strict restrictions on the parameters of (\ref{eq:CNGLE}) except for the very broad ones imposed by Propositions. Moreover, the parametric space of the solution based on this approximation has reduced dimensionality ($C$, $b$, and $\Sigma$\footnote{$\Sigma$-parameter can be considered as irrelevant in some sense because the parametric space topology is defined by $\Sigma=0$ and $\Sigma_{\pm}$-isosurfaces.}).

\begin{figure}
    \centering
    \includegraphics[scale=0.41]{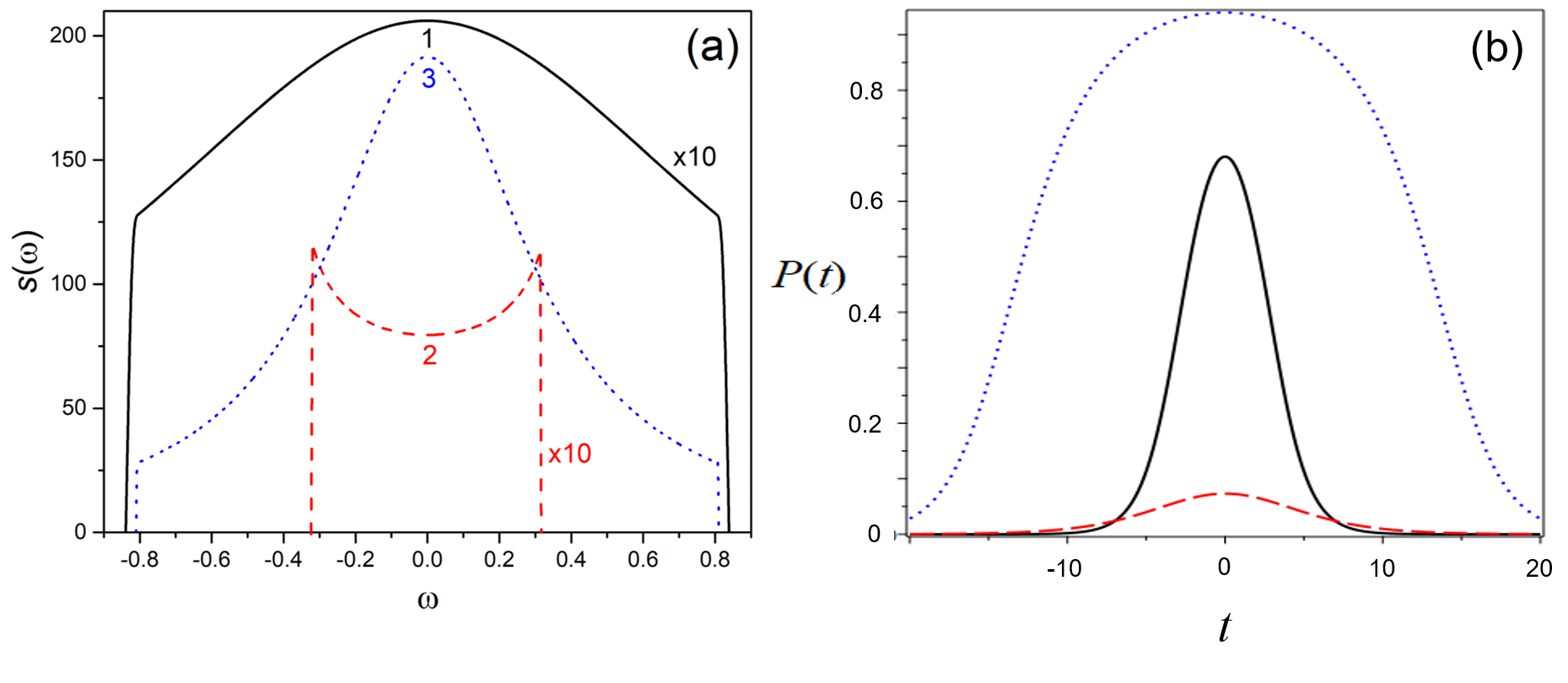}
    \caption{Dimensionless spectra (a) and temporal profiles (b) of DS. Solid black curves (1): $C=1$, $b=20$; Dashed red curves (2): $C=1$, $b=0.2$; Dotted blue curves (3): $C=2/3$, $b=20$. $\Sigma=0.01$. Scale for the black solid and red dashed curves in (a) increases tenfold.}
    \label{fig:fig6}
\end{figure}

The adequateness of the considered approach as well as its compliance with that based on the solution (\ref{eq:ansatz}) (e.g., see \cite{renninger2008dissipative}), in particular, is demonstrated by a ``zoo'' of spectral and temporal DS shapes obtained from Eqs. (\ref{eq:P1},\ref{eq:P2},\ref{eq:tchirp},\ref{eq:spectrum}) (Fig. \ref{fig:fig6}). We consider only ($+$)-branch and $b>0$ (self-enhancing self-phase modulation\footnote{Self-enhancing self-phase modulation could be interpreted, for instance, in the following way. In a Kerr-lens mode-locked laser, the mechanism of ultrashort pulse formation is the loss decrease due to a laser beam self-focusing \cite{brabec1992kerr}. That means  beam squeezing and, thereby, the self-phase modulation growth $\propto w^{-2}$, where $w$ is a beam size.}).  

The main feature of the approach considered above is that it is built in the spectral domain. Therefore, the spectral shapes could be considered as a roadmap to a DS classification\footnote{The example of the experimental implementation of such classification could be seen in \cite{rudenkov2022route}.}. One may see three main types of spectra from Fig. \ref{fig:fig6} (a): convex (1), concave (2), and finger-like (3). The first and third types correspond to a large $b$, i.e., a small contribution of the imaginary quintic term in (\ref{eq:CNGLE}). These spectra relate to Eq. (\ref{eq:RJ}) and, thereby, represent a truncated Lorentzian so that a transition between them is defined by the condition of $\Xi = \Delta$. The transition from (1) to (3) represents a shift from a soliton-like temporal profile to a stretched and flattened one, demonstrating an energy harvesting mechanism due to DS broadening. One has to note that the cutoff frequency remains almost the same in this case. That is the pulse width scales as $\propto 1/\Xi$. As it will be shown below, such transformation of the DS spectrum demonstrates a transition to the DSR.  

When the contribution of the positive imaginary quintic term in (\ref{eq:CNGLE}) grows ($b \to 0$), the spectrum becomes concave (the red dashed curves in Fig. \ref{fig:fig6})\footnote{It should be noted that such concave spectra are not the product of the high-order GDD action, which could be described by an additional term $\propto i \, \partial^4/\partial t^4$ in (\ref{eq:CNGLE}) \cite{kalashnikov2005approaching}.}. In this case, the DS energy (compare black solid and red dashed curves in Fig. \ref{fig:fig6}) decreases. That results from the chirp degradation for a chosen value of $C$ (see Fig. \ref{fig:fig5}). The DS energy could be more significant for a smaller $C$-parameter for the considered case. The energy scaling is provided by the DS stretching but without a profile flattening.

\begin{figure}[h]
  \centering
  \begin{tabular}{c c c}
    \includegraphics[scale=0.19]{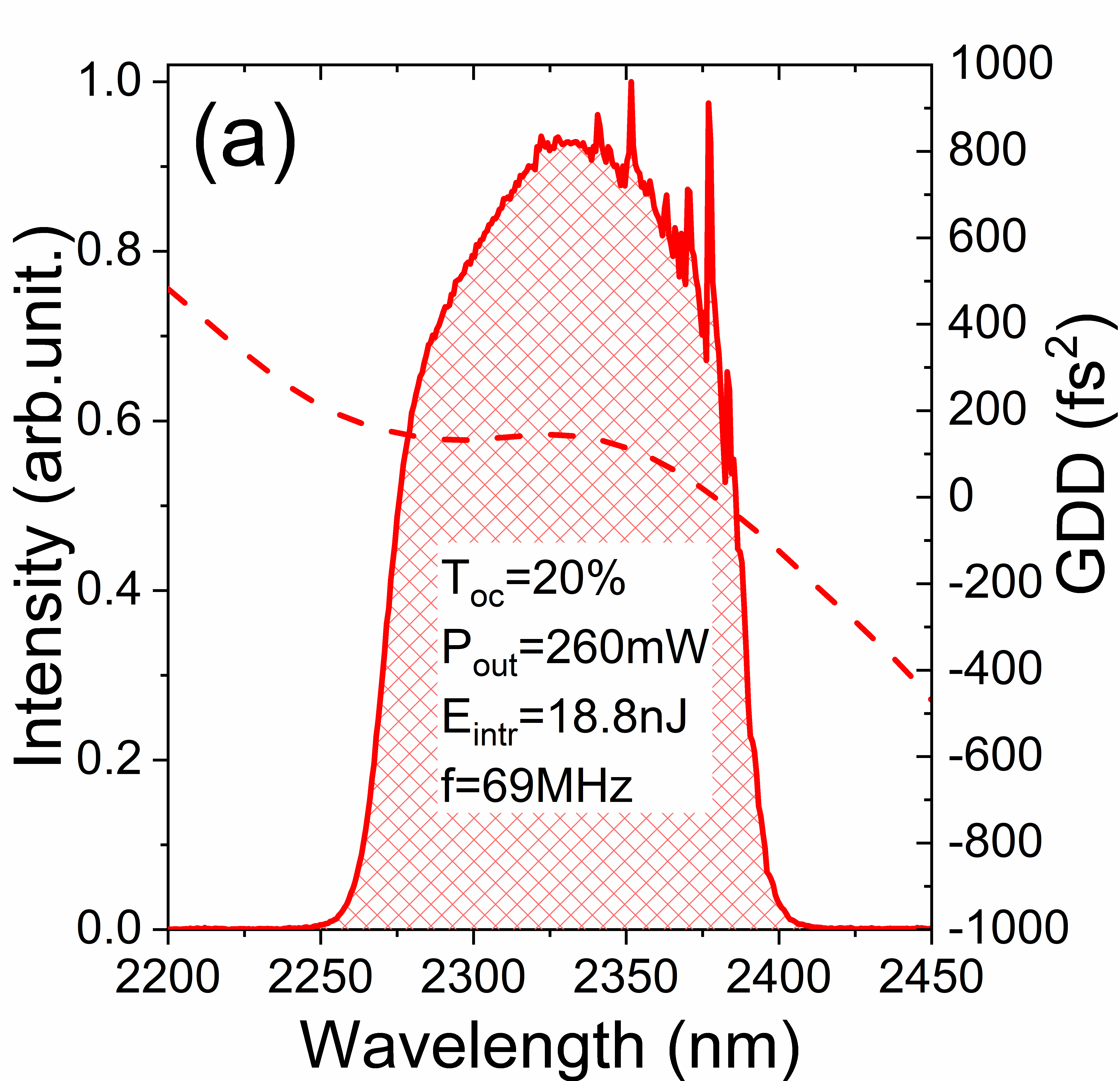} 
    \includegraphics[scale=0.19]{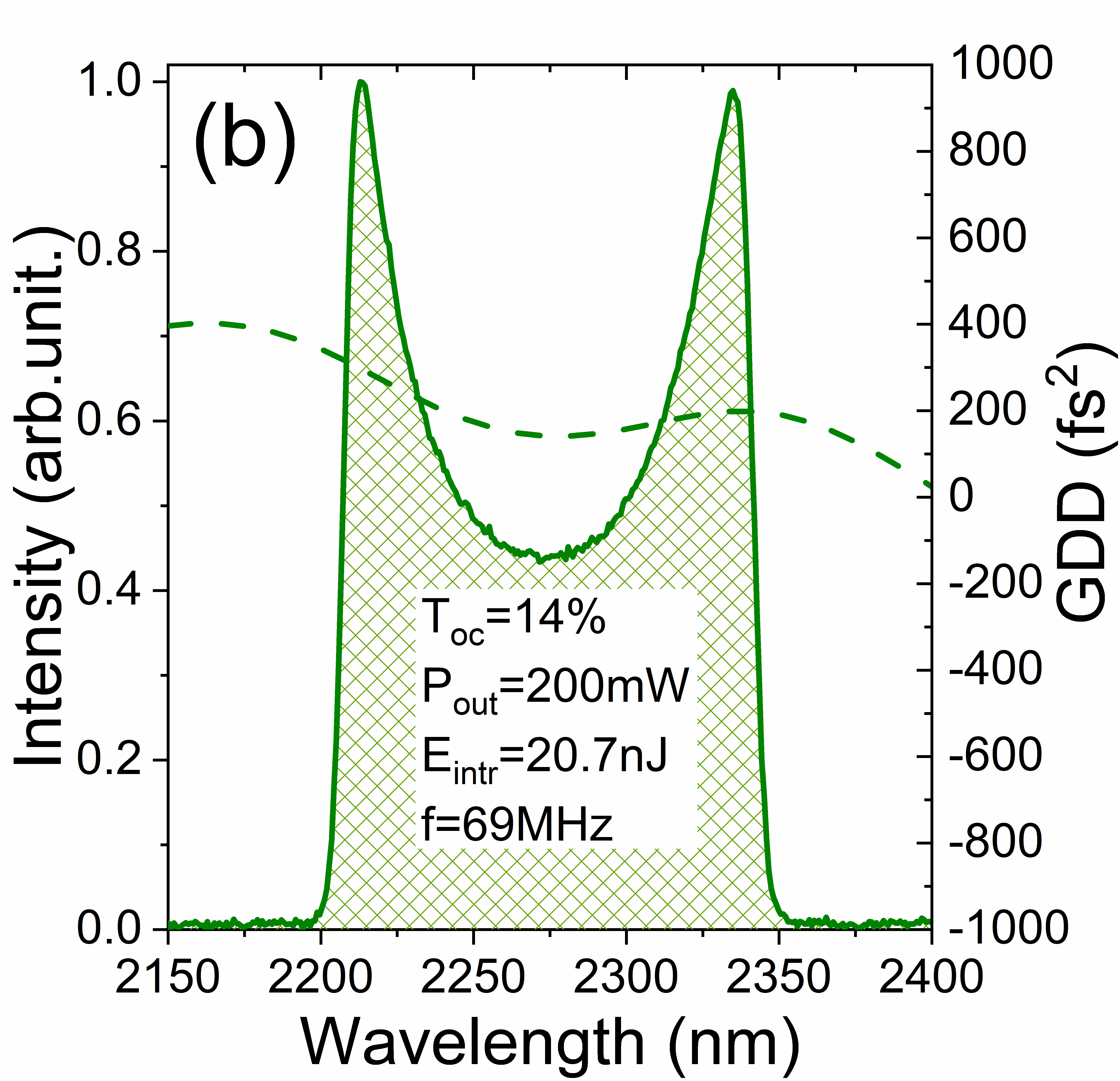}
    \includegraphics[scale=0.19]{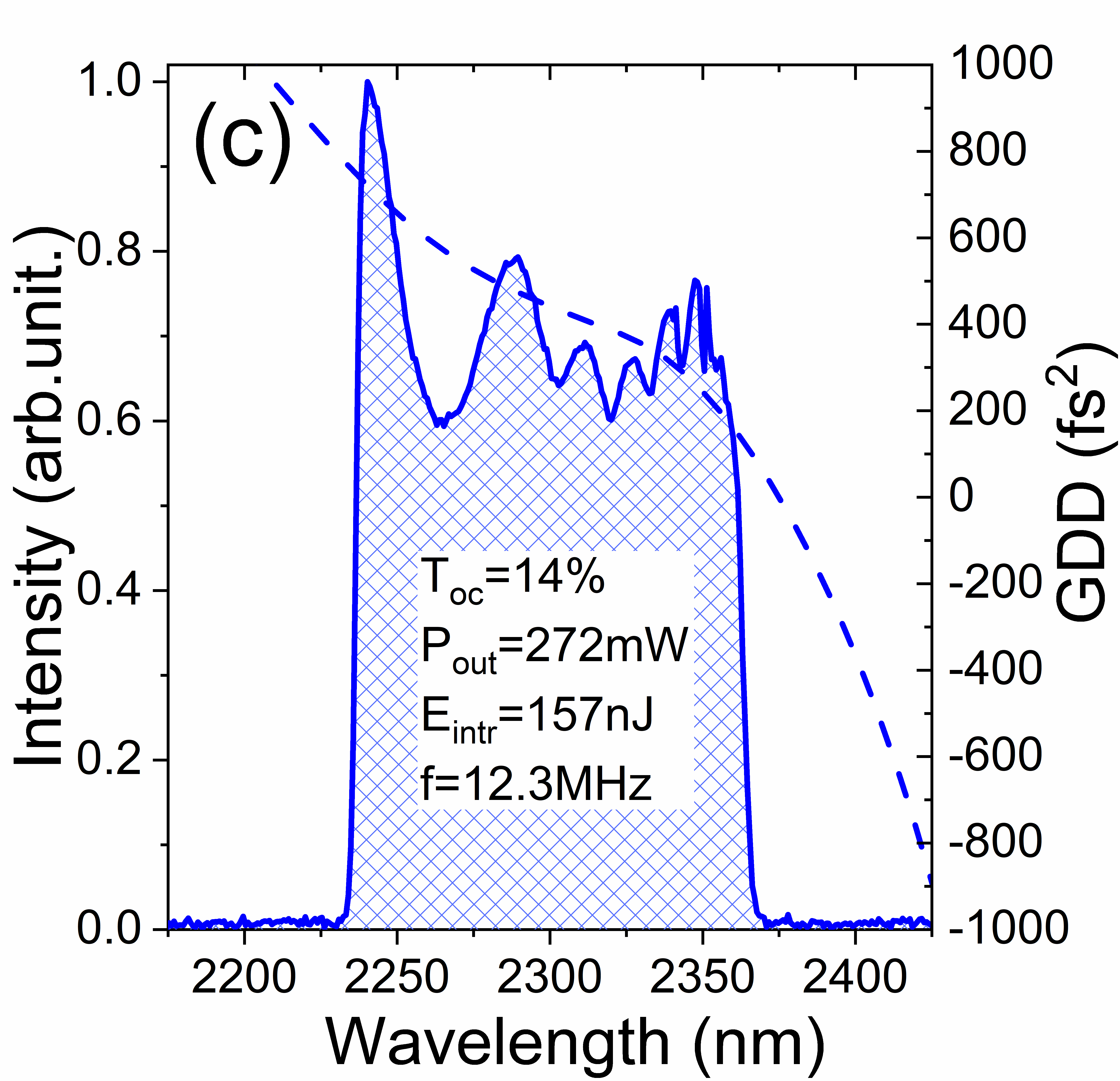}
  \end{tabular}

  \caption{Experimental CPO spectra (solid lines) obtained during pulse energy scaling and corresponding dispersion curves (dashed lines): (a) bell shape spectrum at the threshold of CPO operation, (b) M-shape spectrum obtained with slightly increased average GDD and pump power values \cite{sorokin2022atmospheric}, (c) near finger-like spectrum after energy scaling by pump power and pulse repetition frequency in DSR \cite{rudenkov2022route}. $T_{OC}$, $P_{out}$, $E_{intr}$, and $f$ are the laser output mirror transmission, output power, intracavity energy, and DS repetition rate, respectively.}
    \label{fig:figureexp2}
\end{figure}

Fig. \ref{fig:figureexp2} illustrates the experimentally observed spectral profile evolution during the DS energy scaling in chirped pulse oscillators (CPO) \cite{rudenkov2022route,sorokin2022atmospheric}. Spectra in Figs. \ref{fig:figureexp2} (a) and (b) were obtained in an oscillator capable of generating both Schr\"{o}dinger-like and DS by smooth tuning of the cavity GDD from negative to positive values. Spectrum in Fig. \ref{fig:figureexp2} (a) was obtained near the CPO threshold, while spectrum (b) was obtained with slightly increased average GDD and pump power. Pulse energy scaling was demonstrated in the CPO cavity with reduced pulse repetition frequency $f$ (12.3 compared to 69~MHz), when we further increased the positive cavity GDD and pump power to maintain the DS stability. The slight asymmetry of the spectrum (Fig.\ref{fig:figureexp2} (c)) is associated with an uncompensated third-order GDD. The narrowband spectral features result from the water vapour absorption in the atmosphere \cite{kalashnikov2011chirped}.


It should also be noted that the adiabatic approximation provides an adequate description of DS spectra even beyond the validity of Proposition 1. Namely, the spectra transform from concave to concave-convex when $\kappa >\gamma$ and $\chi \neq 0$, as it was described in \cite{kalashnikov2009chirped,akhmediev2008roadmap}. Moreover, there are classes of unusual DS solutions for $b <0$, for instance, spike on a background or parabolic-like.\footnote{The parabolic-like pulses have $d\Omega/dt \to \pm \infty$ on the edges, i.e., such DS is truncated on $t$.}

\subsection{DS compressibility and its fidelity}\label{fid}

Eq. (\ref{eq:sprofile}) provides information about the DS internal phase profile. This profile is inhomogeneous, which troubles its compression. Such compression would allow producing, for instance, high-intensive ultrashort laser pulses for numerous applications.  

For simplicity, let us assume that $\chi=0$. Returning to the dimensional values, one may wright\footnote{Eq. (\ref{eq:sp}) demonstrates that the chirp is proportional to $\gamma^2/\kappa \zeta$, which clarifies Proposition 1.}    

\begin{equation} \label{eq:sp}
    e(\omega)=\sqrt{\frac{6\pi\gamma}{\zeta  \kappa}} \frac{{\mathrm e}^{\frac{\frac{3}{2} \mathrm{i}  \gamma^{2} \omega^{2}}{\beta  \kappa  \zeta  \left(\Xi^{2}+\omega^{2}\right) \left(\Delta^2 -\omega^2 \right)}} }{\sqrt{  \mathrm{i}\left(\Xi^{2}+\omega^{2}\right) }}\mathcal{H}\left(\Delta^2-\omega^2\right).
\end{equation}

DS would be maximally compressible if its spectral phase $\varphi(\omega)= \Upsilon \times \omega^2$ ($\Upsilon$ is a spectral chirp). Such a phase could be externally compensated by an appropriate GDD $\beta= -\Upsilon$ that would lead to temporal ``focusing'' of DS in agreement with the principle of space-time duality in optics \cite{kolner1994space}. However, Eq. (\ref{eq:sp}) demonstrates that the spectral phase of DS is not purely quadratic in $\omega$, that is, the spectral chirp $\Upsilon$ is frequency dependent. However, the phase distortion is maximally suppressed or ``flat'', when $\Xi = \Delta$: $\Upsilon \propto (\Delta^4-\omega^4)^{-1}$. Such a ``flatness'' allows a DS compression with minimal fragmentation, or maximal ``\emph{fidelity}''.

Let us return to the dimensionless values in (\ref{eq:sp}) so that the amplitude $e(\omega)$ is normalized to $\sqrt{\kappa/\beta}$, $d=\gamma/\kappa$ and the frequencies are normalized as above. Then, we perform the same procedure as before based on the stationary phase method by expanding the phase into the Taylor series and keeping the term $\propto \omega^2$ in the integral $\int_{-\infty }^{\infty } e(\omega')\exp(i\omega't)d\omega'$. The corresponding term is

\begin{equation}\label{eq:cd}
    \Upsilon'=\frac{6 d \Delta'^2 \Xi'^2+\pi(\Delta'^2-\Xi'^2)}{4\Delta'^4\Xi'^4},
\end{equation}
\noindent which gives a value of group-velocity dispersion required for the DS compression.

\section{Master diagram and dissipative soliton resonance} \label{dsr}

Eqs. (\ref{eq:spectrum},\ref{eq:RJ}) allow finding the DS energy: $E=\frac{1}{2\pi}\int_{-\infty }^{\infty }s(\omega)d\omega$. This integral can be evaluated numerically in the general case or found in the closed form for $\chi=0$:

\begin{equation}\label{eq:en}
   E=\frac{6 \gamma \arctan\! \left(\frac{\Delta}{\Xi}\right)}{\zeta \kappa \Xi}.
\end{equation}
 At this stage, we can introduce the following definition:
 \begin{definition}[\textbf{Master diagram}]
The master diagram represents a DS parametric space in ($C-E$)-coordinates. 
\end{definition}
The DS parametric space represented by the master diagram is confined by a vacuum instability threshold $\Sigma=0$ and filled by ``isogains'' $\Sigma=const>0$ which extreme points $\frac{d C}{dE}=0$ define the curve dividing ($\pm$)-DS branches.
All DS parameters are implied to be dimensionless (e.g., $E$ in (\ref{eq:en}) could be normalized to $\kappa \sqrt{\zeta/\beta \gamma}$ for the above frequency normalization).
The diagram can contain other physically sound curves (e.g., ``fidelity curve'' $\Xi=\Delta$) and regions. Moreover, the ``web'' of isogains is deformable by finite $b$, and the disjointed islands (e.g., for $b \lessgtr0$) may coexist. In the latter case, one should be cautious and check the physicality and stability of solutions.

\begin{figure}
    \centering
    \includegraphics[scale=0.35]{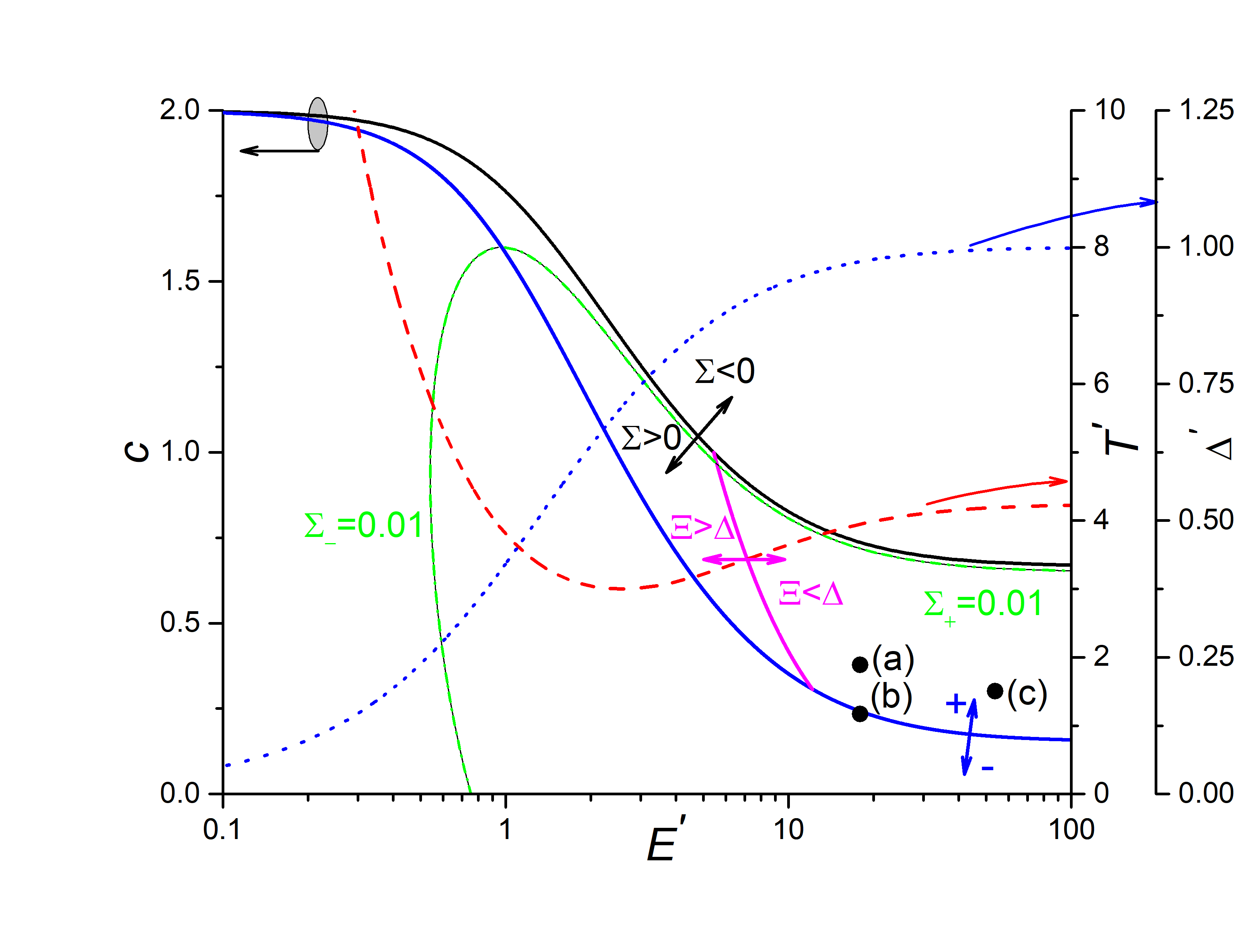}
    \caption{Master diagram for $\chi=0$. Curves related to the left vertical axis: vacuum instability border $\Sigma=0$ (solid black); the division between ($\pm$)-branches of DS solutions (\ref{eq:div}) (solid blue); ``fidelity curve'' (solid magenta); one of the isogains ($\Sigma_{\pm}=0.01$, dashed-dotted green). The right vertical axes relate to the DS temporal $T'$ (red dashed curve) and spectral (dotted blue curve) $\Delta'$ widths along the stability border $\Sigma=0$, respectively.  $C=\alpha \gamma/\beta \kappa$, $E'=E \times \kappa \sqrt{\zeta/\beta \gamma}$, $\Delta'=\Delta \times \sqrt{\beta \zeta/\gamma}$, $T'=T \times (\kappa/\sqrt{\gamma \zeta \beta})$. Points (a), (b), and (c) refer to Figs. \ref{fig:fig12},\ref{fig:fig13}.}
    \label{fig:fig8}
\end{figure}

\begin{figure}
    \centering
    \includegraphics[scale=0.35]{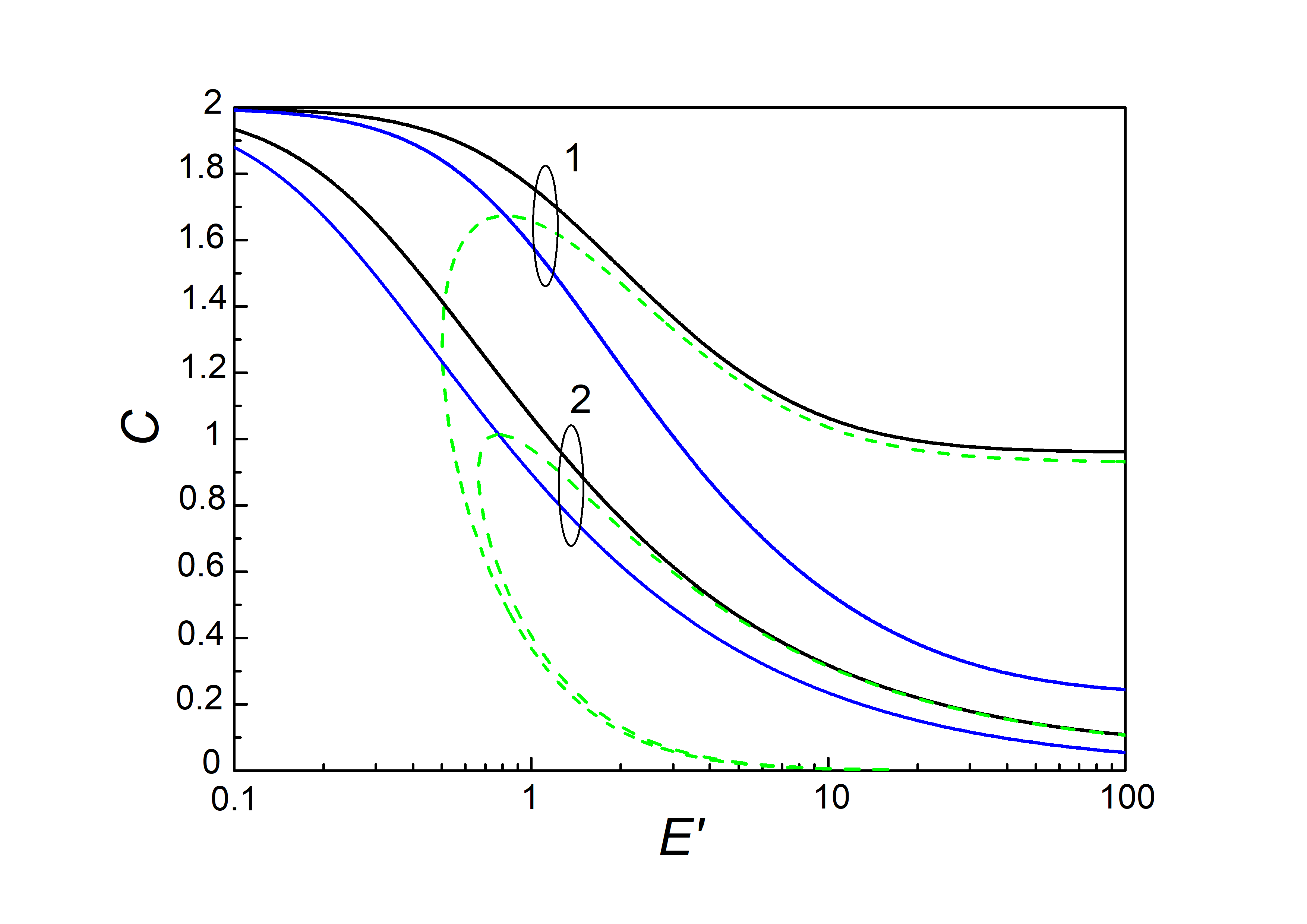}
    \caption{Master diagram for $\chi=$-5 (1), 0.2 (2). Black solid curves are the vacuum instability thresholds, blue solid curves divide ($\pm$)-branches of DS solutions, and green dashed lines correspond to isogains $\Sigma=0.01$. One can see as a region of DSR squeezes and shifts to the smaller $C$ for $b=0.2$ in parallel with the corresponding chirp transformation (Fig. \ref{fig:fig5}).}
    \label{fig:fig9}
\end{figure}

The master diagrams for $b\gg 1$, $b=$0.2, and -5 are shown in Figs. \ref{fig:fig8}, \ref{fig:fig9} \cite{kalashnikov2009chirped,rudenkov2022route}. This diagram allows formulating the notion of DSR \cite{chang2008dissipative}.

 \begin{definition}[\textbf{Dissipative soliton resonance}]
$\exists \,C^*:\lim_{C \to C^*}E=\infty$ or there exists a set of $C$-parameters providing an infinite energy asymptotic\footnote{The term of DSR was invented in \cite{chang2008dissipative} based on the method of moments. The variational method leads to a softer definition: there is asymptotics $E \propto C^{-p}$, where $p=1/2$ for CQGLE with $\chi=0$ \cite{kalashnikov2018theory}. The variational approximation represents DSR as a range of master diagram.}.  
\end{definition}

One can see from Fig. \ref{fig:fig8} that ($+$)-branch of DS is energy-scalable in the sense of Definition 2. The bottom border of the corresponding region is characterized by:

\begin{equation}\label{eq:res}
    \begin{matrix}
E=\frac{6 \sqrt{2\gamma \beta}}{\kappa \sqrt{\eta}}\frac{\arctan(\frac{\sqrt{3}\sqrt[4]{\Sigma}}{\sqrt{6-13\sqrt{\Sigma}}})}{\sqrt{6-13\sqrt{\Sigma}}}, \\
C=2-4\sqrt{\Sigma}, \\
P_0=\frac{3\sqrt{\Sigma}}{2 \zeta}, \\
\Delta^2= \frac{3\gamma\sqrt{\Sigma}}{2\beta \zeta},\\
\Xi^2=\frac{\gamma}{2\zeta \beta}(6-13\sqrt{\Sigma})
\end{matrix}
\end{equation}

\noindent within $\mathbf{\Sigma \in \left[ 0,36/169 \right]}$. Thus, the DSRs $E \to \infty$ are located between $\mathbf{C=2/3, \,2/13}$\footnote{See Proposition 2, where we limited ourselves by the interval approximately corresponding to DSR.}. On the vacuum instability border $\Sigma=0$, one has:

\begin{equation}\label{eq:res2}
    \begin{matrix}
E \to \infty,\\
P_0 \to \zeta^{-1}, \\
 \Delta^2\to\gamma/\beta \zeta,\\
\Xi\to0.
\end{matrix}
\end{equation}
\noindent and $\mathbf{C=2/3}$. Such a divergence of energy for a given parameter $C$ and $\Sigma$ lying within the interval defined above can be considered as the \textbf{DSR definition} \cite{chang2008dissipative}. For a laser oscillator, the DS energy could be scaled not only by pump power or/and laser beam cross-section (within a limited range) but also mainly by the oscillator period scaling up to times close to the gain relaxation time. While all other parameters, such as GDD ($\beta$) in Eq. (\ref{eq:CNGLE}), spectral dissipation ($\alpha$), and nonlinear parameters $\gamma$, $\kappa$, and $\xi$ remain unchanged.

Eqs. (\ref{eq:res},\ref{eq:res2}) demonstrate important signatures of transition to DSR: cut-off frequency $\Delta$ (DS spectrum half-width) tends to a constant; ``chemical potential'' $\Xi$ tends to zero (slectrum becomes ``finger-like''), and a peak power becomes above-confined. Owing to the latter, the DS width scales with energy. The DS width can be estimated from Eq. (\ref{eq:tchirp}) by its integration that gives for $\chi=0$ the implicit DS temporal profile:

\begin{equation}
    t=\frac{3 \gamma^{2} \left(\frac{\arctan\left(\frac{\Omega \left(t\right)}{\Xi}\right) \Delta}{\Xi}+\mathrm{arctanh}\! \left(\frac{\Omega \left(t\right)}{\Delta}\right)\right)^{2}}{\beta  \zeta  \kappa  \Delta  \left(\Delta^{2}+\Xi^{2}\right)},
\end{equation}

\noindent where $\Omega(t)$ and $P(t)$ are connected through Eq. (\ref{eq:P1}). Then, the DS width can be expressed as $T=\frac{3 \gamma^{2}}{\beta  \zeta  \kappa  \Delta  \left(\Delta^{2}+\Xi^{2}\right)}$.

These tendencies are illustrated in Fig. \ref{fig:fig8}. One may see that the transition to DSR, with the subsequent change of a DS shortening by its broadening and simultaneous ``freezing'' of spectral width growth, is accompanied by a crossing of the maximum fidelity curve. The latter means a visible growth of spectral peak, i.e., transition to a ``finger-like'' spectrum.

\begin{figure}
    \centering
    \includegraphics[scale=0.37]{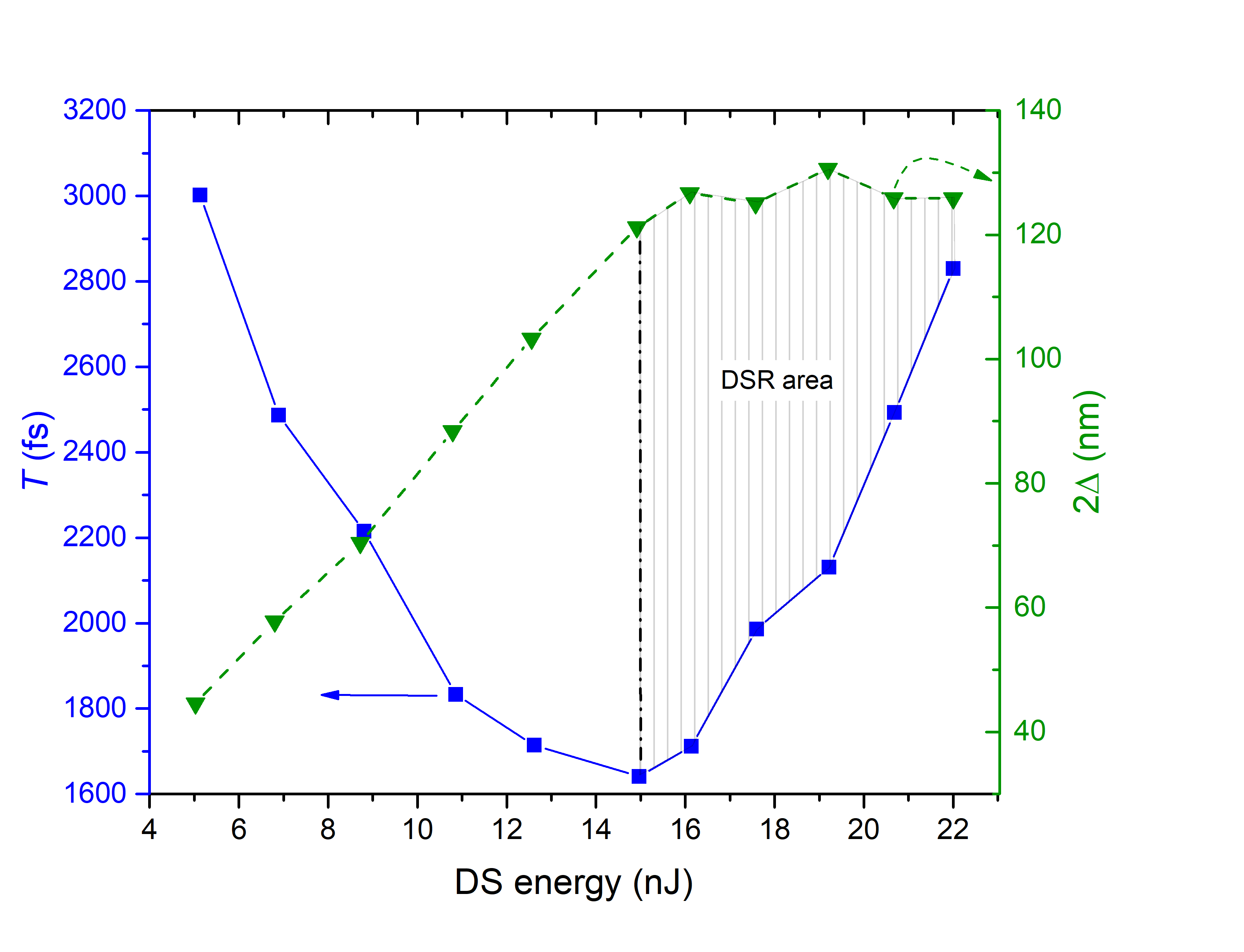}
    \caption{Experimental master diagram \cite{rudenkov2022route} demonstrating a transit to DSR regime via asymptotically constant DS spectral width $\Delta$ and its temporal width $T$ scaling. A spectrum becomes finger-like.}
    \label{fig:figureexp1}
\end{figure}

All these manifestations are experimentally verifiable. Fig. \ref{fig:figureexp1} shows the experimental master diagram obtained in the experiments with Cr:ZnS chirped-pulse oscillator \cite{rudenkov2022route}. We can interpret this diagram as a manifestation of DSR owing to the following facts: i) spectral with becomes asymptotically constant and ii) DS width growth asymptotically with energy. The transition point is a change of the DS width decrease (energy scaling by peak power growth) by the DS width increase (scaling by the DS stretching). It is very characteristic that a ``finger'' appears near the DS spectrum center (Fig. \ref{fig:figureexp2}, c) that results from the $\Xi$-decrease.

\section{DS thermodynamics}\label{sec10}

Perhaps the most exciting advance of the approach considered above is that its main results are formulated in the spectral domain (``momentum space''). That allows applying the notions of kinetic theory straightforwardly to DS so that the latter could be understood in terms of an incoherent/semicoherent condensate of incoherent nonlinear waves \cite{picozzi2009thermalization}. 

\begin{figure}
    \centering
    \includegraphics[scale=0.8]{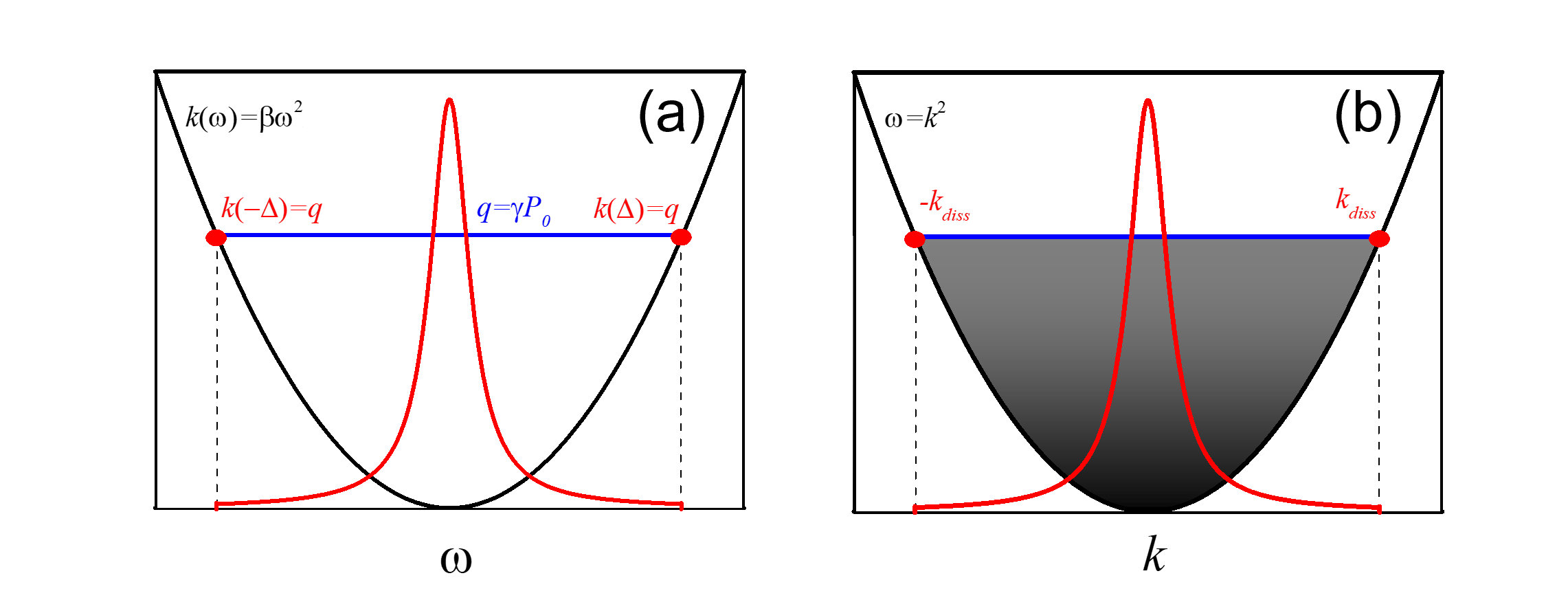}
    \caption{(a): DS spectrum (\ref{eq:RJ}) (red Lorentzian curve) and a wave-number of linear waves (black parabolic curve), which resonance with DS is denoted by red points. (b): The turbulence spectrum in the wave-number space (red curve) and the Langmuir dispersion curve (black). See main text for the comments.}
    \label{fig:fig11}
\end{figure}

Let us limit ourselves to the case of $\chi=0$. Eq. (\ref{eq:RJ}) demonstrates the well-known Rayleigh-Jeans equilibrium distribution \cite{zakharov2012kolmogorov} with a negative ``chemical potential''  $-\mu=\Xi^2$ and a ``temperature'' $\Theta=6\pi \gamma/\kappa \zeta$. This spectrum and its counterpart from the turbulence theory \cite{robinson1997nonlinear} are shown by red curves in Fig. \ref{fig:fig11}. The Lorentzian turbulence spectrum results from the $k \to 0$ condensation of waves with the Langmuir dispersion relation $k=\omega^2$, as shown by a graded shading in Fig. \ref{fig:fig11} (b).  

The DS wave number is $q = \gamma P_0= \beta \Delta^2$ from Eq. (\ref{eq:cutoff}), which is analog to the soliton area theorem $q = \gamma P_0/2 =\beta T^2/2$ \cite{renninger2010area}. The equality of the DS wave number $q=\gamma P_0$ with that of linear waves $k(\omega)=\beta \omega^2$ (compare with the Langmuir dispersion curve in Fig. \ref{fig:fig9} (b)) defines a DS spectral (half-)width $\Delta$. In the case of turbulence, a wave number cut-off is provided by dissipation, but dissipation is also a vital factor for DS. Roughly from Eq. (\ref{eq:CNGLE}), the spectral dissipation $~\alpha \Delta^2$ has to be compensated by a nonlinear gain $~\kappa P_0$ ($\Sigma=0$ on the vacuum instability border)\footnote{Spectral filtering causes a cut-off on the pulse edges, where the spectral components with maximal frequency deviation are located \cite{bale2008spectral}. Simultaneously, a nonlinear gain is roughly defined by a DS peak power, which is usually concentrated towards a DS center $t=0$ or is constant for a flat-top pulse in the DSR regime. Such a gain provides a spectral loss compensation on the DS edges due to energy redistribution inside a chirped DS \cite{ankiewicz2008dissipative}.}. Hence, a combination of the dispersion/dissipation balances leads to $\alpha \gamma/\beta \kappa \simeq 1$, or \emph{soliton condition} (``potential condition'') implying a Gibbs-like steady-state probability distribution in statistical mechanics \cite{katz2006non} (Proposition 2)\footnote{It should be noted, that $\beta-$sign is opposite to that in the nonlinear Schr\"{o}dinger equation, therefore this condition is soliton-\emph{like}.}. 

As the physical explanation of the DS formation, one may say additionally to the previous discourse the following. The stability of a DS in conditions of strong non-equilibrium is maintained through efficient energy exchange with the surroundings and internal energy redistribution. This process generates an energy flow within the DS, leading to phase inhomogeneity. The spectral broadening, caused by DS chirping, transforms spectral dissipation at the DS spectrum edges into a structured energy exchange: energy inflows into the pulse center, where the spectral losses are minimal and a nonlinear gain compensates losses (both linear and spectral). But at the DS wings, where the spectral deviation is maximal due to chirp, energy dissipates. A crucial aspect of this process is the chirp. Without the chirp, spectral dissipation uniformly affects the pulse, potentially causing multi-pulse instability \cite{kalashnikov2003multipulse}. Conversely, a chirp that varies with power creates an inhomogeneity in energy transfer, directing energy flow from areas near the central wavelength—where the gain is highest—toward the pulse's wings, where energy is then dissipated, thereby the pulse's localization is maintained through spectral dissipation via a non-linear chirping mechanism. That introduces an additional stabilization mechanism through saturable self-amplitude modulation, significantly enhancing the DS's robustness across a wide range of laser parameters \cite{bale2008spectral}.

These observations on the DS properties testify about an immediate relation between DS and a family of incoherent/semicoherent solitons \cite{picozzi1,picozzi2,rotschild2008incoherent}. That means that the DS thermodynamics has to be based not only on considering the DS interaction with an external thermal basin but on a view of DS as a microcanonical statistical ensemble of independent ``\emph{quasi-particles}'' confined by a collective potential (\ref{eq:RJ}) \cite{picozzi1}. 

One may indirectly test this proposition through a numerical experiment. For this goal, we have to take into account the energy dependence of $\sigma$-parameter in Eq. (\ref{eq:CNGLE}) assuming that it describes a saturable net-loss in a laser \cite{kalashnikov2009chirped}: $\sigma \approx \delta (E/E^*-1)$, where $\delta \equiv dE/dE^*|_{E=E^*}$, and $E^*$ is the energy of continuum wave generation at $\sigma=0$ (now, its normalized value replaces $E'$ in Fig. \ref{fig:fig8}). Also, we include the thermal basin, which is described as an additive complex noise term $\Gamma$ in Eq. (\ref{eq:CNGLE}). It is assumed to be Gaussian and uncorrelated:

\begin{gather}
  \nonumber  \left\langle \Gamma(z_1,t_1) \Gamma^*(z_2,t_2)\right\rangle=\Theta_b\delta(z_1-z_2)\delta(t_1-t_2),\\
\left\langle \Gamma(z_1,t_1) \Gamma(z_2,t_2)\right\rangle=0,
\end{gather}   
\noindent where $\Theta_b$ is noise's spectral power (temperature).

\begin{figure}[h]
  \centering
  \begin{tabular}{@{}c@{}}
    \includegraphics[scale=0.4]{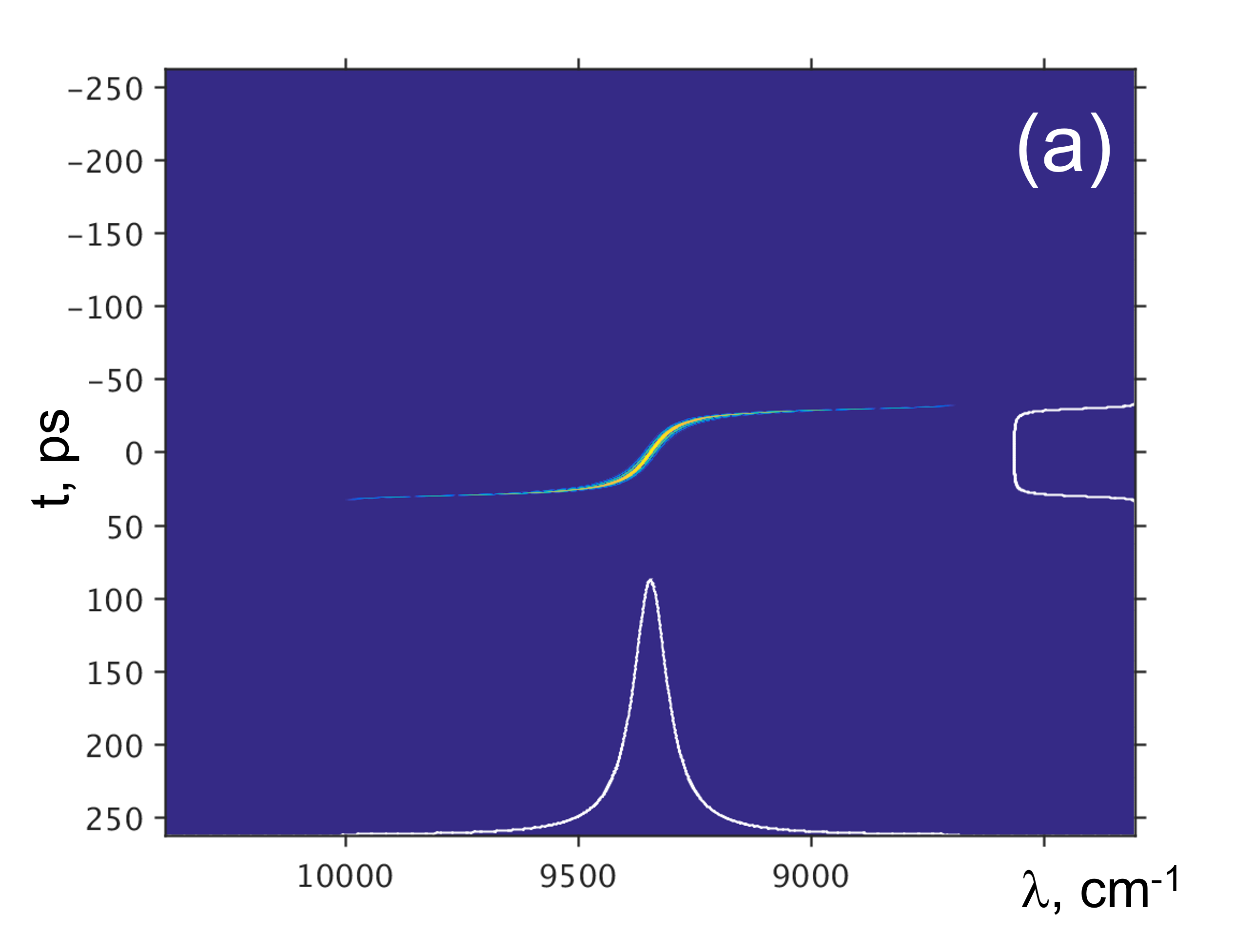} \\[\abovecaptionskip]
  \end{tabular}


  \begin{tabular}{@{}c@{}}
    \includegraphics[scale=0.4]{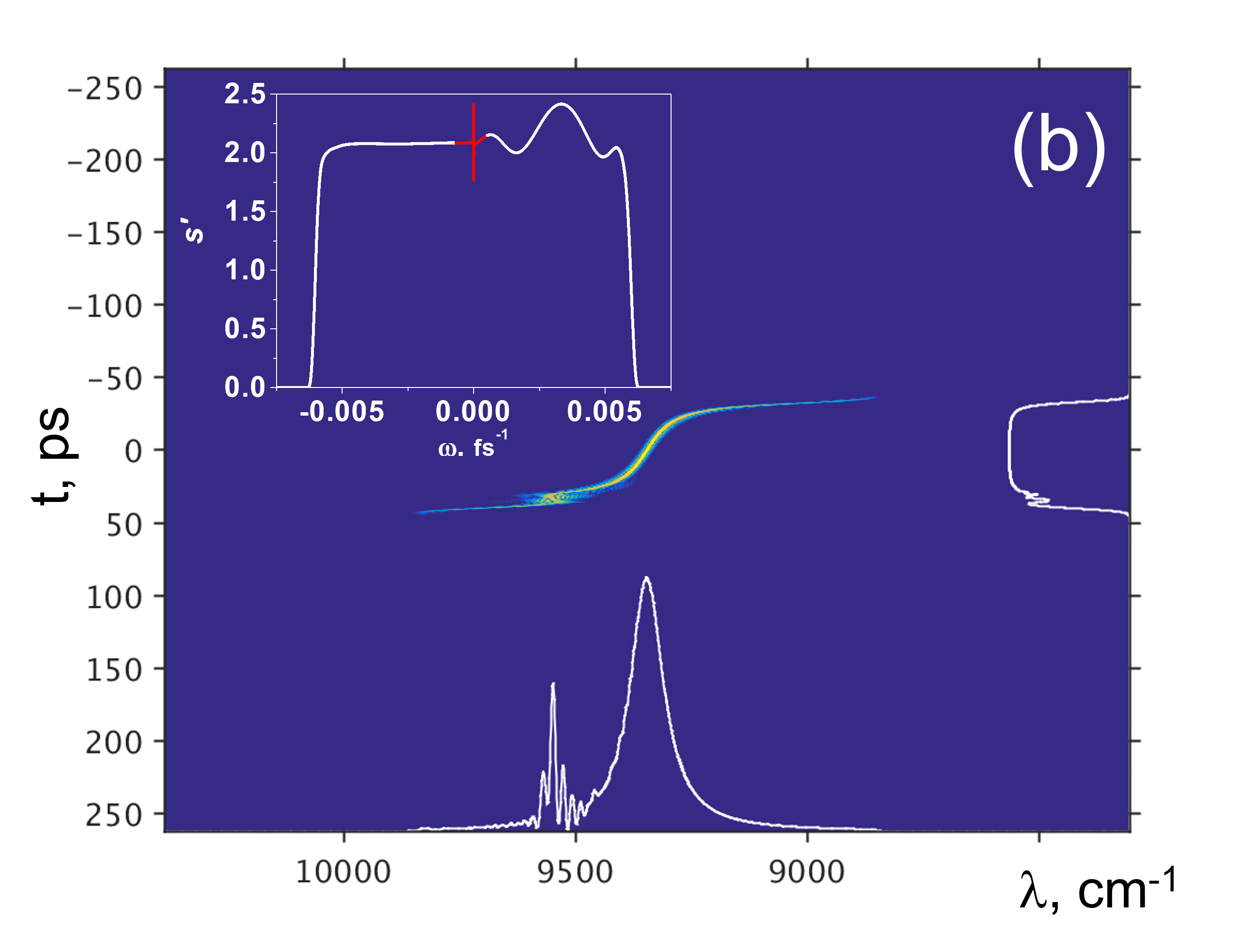} \\[\abovecaptionskip]
  \end{tabular}
\caption{DS spectra (bottom axis), temporal profiles (right axis), and the corresponding Wigner function \cite{dragoman1997wigner} (center) for $E^*=$18 and $C=$0.24 (a) and 0.18 (b). $\delta=$0.05, $\Theta_b=10^{-10}\gamma^{-1}$, $\chi=$0. Inset - DS spectrum distorted by an absorption line with the dimensionless amplitude 0.0025 and the width of 1GHz \cite{kalashnikov2011chirped}.}
\label{fig:fig12}
\end{figure}

Let's ``wander'' inside a DS master diagram searching for transit to turbulence. The starting point ($a$) (Fig. \ref{fig:fig8}) corresponds to a DSR region with a finger-like spectrum and table-top temporal profile (Fig.  \ref{fig:fig12} (a)). The shift to an area of ($-$)-DS branch (point ($b$) (Fig. \ref{fig:fig8})) excites (slightly decouples) an ``internal modes'' or quasi-particle complexes that manifests itself as a distortion of both spectral and temporal profiles (Fig. \ref{fig:fig12} (b)). As a rule, such distortions are asymmetrical but preserve the DS spectral-time integrity. An inro-DS excitation is illustrated by the inset in Fig. \ref{fig:fig12} (b), where a narrow-band Lorentzian absorption line at the DS spectrum center excites a long-range asymmetric perturbation ``wave'' confined in a collective potential between the perturbation and the DS spectral edge.    

\begin{figure}[h]
  \centering
  \begin{tabular}{@{}c@{}}
    \includegraphics[scale=0.4]{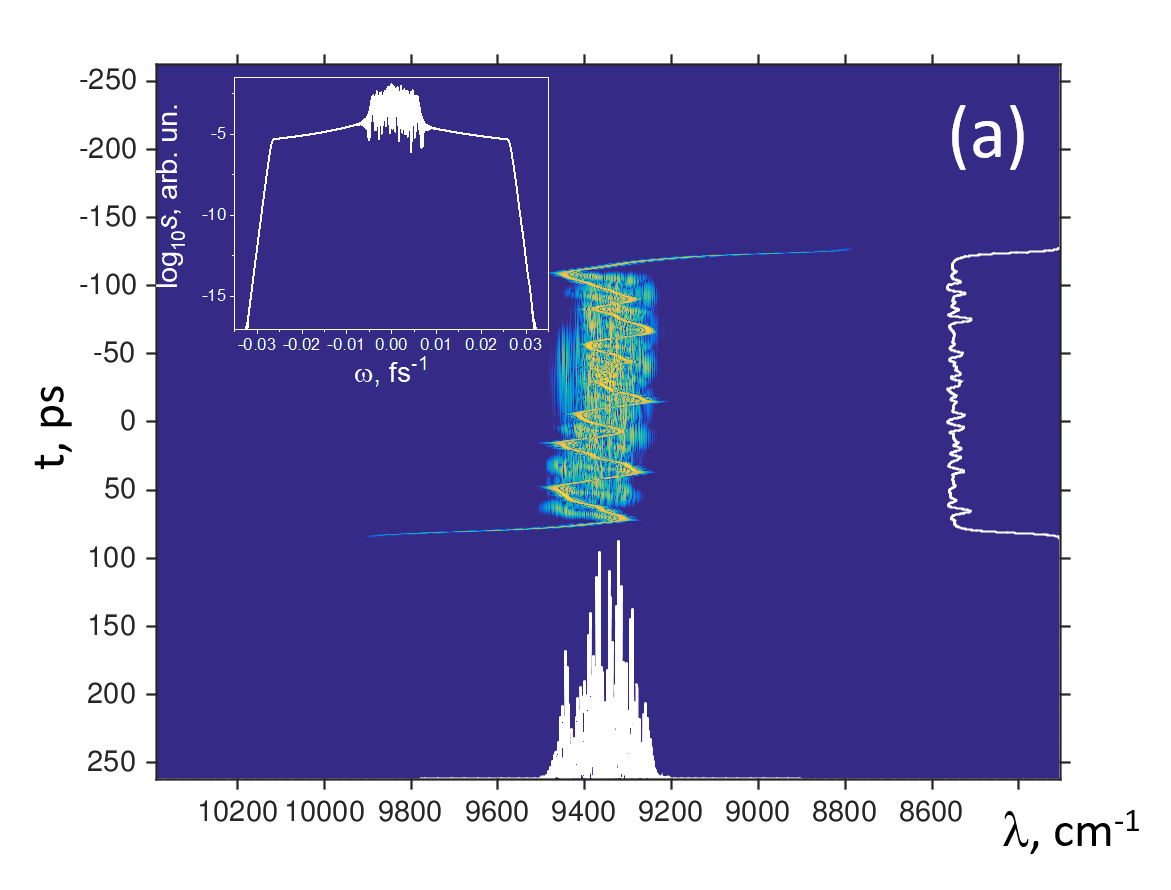} \\[\abovecaptionskip]
  \end{tabular}


  \begin{tabular}{@{}c@{}}
    \includegraphics[scale=0.6]{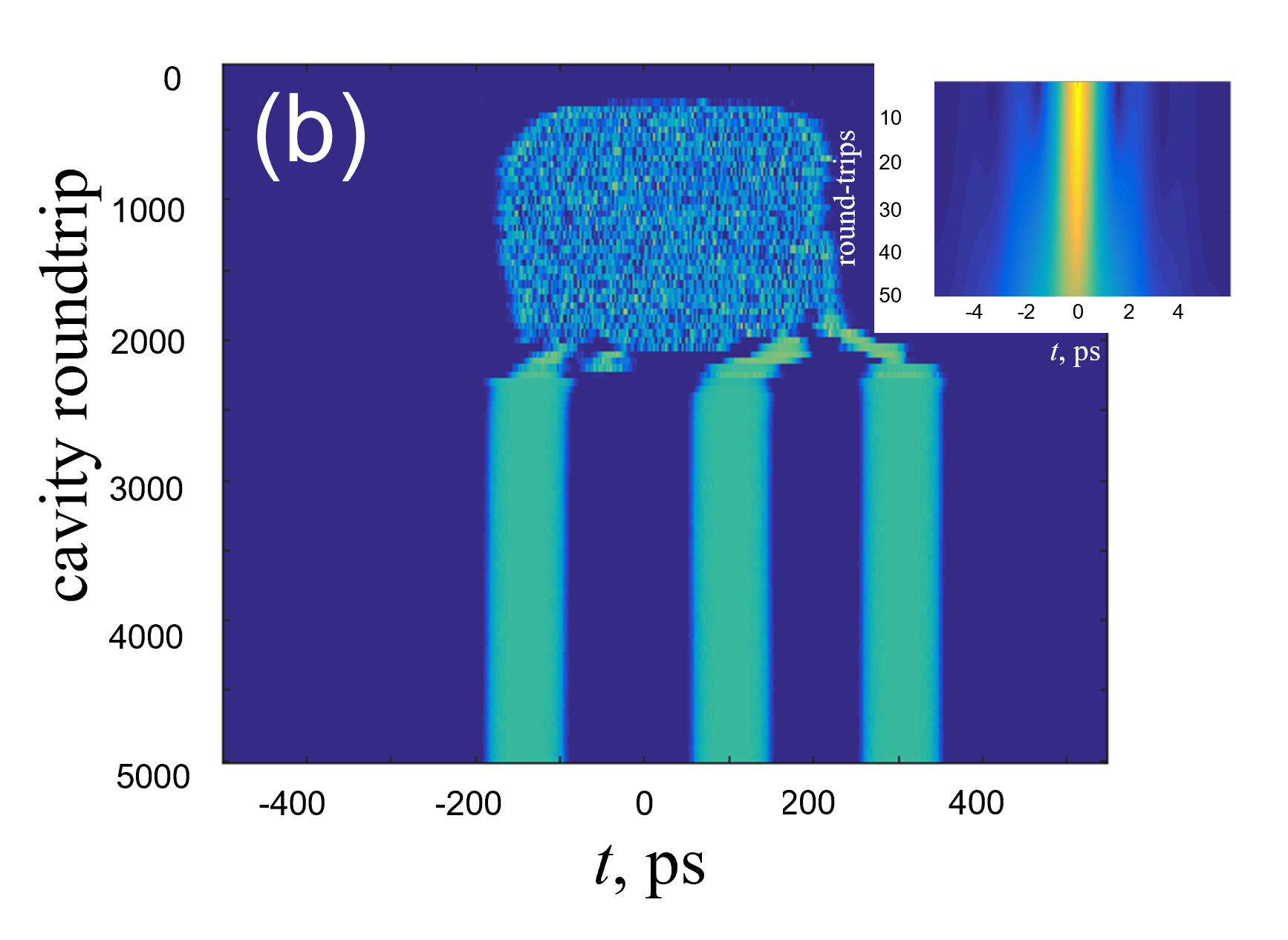} \\[\abovecaptionskip]
  \end{tabular}
\caption{(a) DS spectra (bottom axis), temporal profiles (right axis), and the corresponding Wigner function (center) for $E^*=$54 and $C=$0.19 (point (c) in Fig. \ref{fig:fig8}). Other parameters as in Fig. \ref{fig:fig12}. (b) Multipulsing from turbulence. Laser cavity roundtrip equals $z$ in (\ref{eq:CNGLE}). Left inset in (a): logarithm of spectral power, right inset in (b): a field autocorrelation function.}
\label{fig:fig13}
\end{figure}

The energy growth leads to turbulence (point (c) in Fig. \ref{fig:fig8} and Fig. \ref{fig:fig13} (a)), which is characterized by a Rayleigh-Jeans spectrum (Fig. \ref{fig:fig13} (a); inset) and a localization in both spectral and temporal domains. The Wigner function makes it evident that there are two correlation times: a correlation time of wave in equilibrium defining a confinement potential (a ``homogeneity scale'')  $\Lambda \propto 1/\sqrt{\Xi}$ (``tails'' of the Wigner functions and the DS profile in Fig. \ref{fig:fig13} (a)), and an ``internal'' correlation time (an ``inhomogeneity scale'') $\ell \propto 1/\sqrt{\Delta}$ (a thick ``snake'' in the central part of the Wigner function in Fig. \ref{fig:fig13} (a)). The easily visible ``trajectories'' in the Wigner function center can be interpreted as a visualization of a DS energy in/out-flow induced by Kolmogorow’s turbulence cascade \cite{robinson1997nonlinear}. The existence of internal coherence scale $\ell$ (inset in Fig. \ref{fig:fig13} (b)) can stimulate a spontaneous creation of the coherent DSs from a localized incoherent DS (Fig. \ref{fig:fig13} (b)). Thus, the treatment of DS as a ``quasi-particles ensemble'' could be considered reasonable when $\ell \ll \Lambda$ inside a DSR region \cite{picozzi2009thermalization}. 


Thus, we can base on the following 
\begin{proposition}
    In a DSR region with $\Xi<\Delta$, DS can be considered as a microcanonical ensemble of quasi-particles confined by a collective potential, 
\end{proposition}

\noindent so that the analytical technique formulated above in the spectral domain allows the formulation of the essential DS thermodynamic characteristics \cite{wu2019thermodynamic}. 

Let us assume $\chi=0$. Then from Eq. (\ref{eq:RJ}):

\begin{definition}
    DS temperature $\Theta \equiv 6 \pi \gamma/\zeta \kappa$.
\end{definition}

The DS ``temperature''\footnote{It has a dimensionality of power in our case. Therefore, the dimensionless temperature $\Theta'=6 \pi \gamma/\kappa$ if power is normalized to $\zeta$. We will consider the dimensionless quantities below.} has a sound physical sense. (i) It rises with $\gamma$, i.e., with a chirp. Physically, it means decreasing phase inhomogeneity or a growing tendency to the quasi-particles decoupling. (ii) A temperature increases with the $\kappa \zeta$ decrease. That is, when saturation of self-amplitude modulation vanishes (Eq. (\ref{eq:CNGLE})), the power becomes less confined from the top. As a result, inhomogeneity grows. In both cases, DS ``warms up''.

\begin{definition}
    Chemical potential $-\mu=\Xi^2$.
\end{definition}
From Eq. (\ref{eq:res2}), the chemical potential tends to zero for DSR that corresponds to $E \to \infty$ by analogy with the Bose-Einstein condensation. The field concentrates at $\omega \to 0$ ($s \propto 1/\omega^2$ in an equilibrium) so that a DS (``condensate'') tends to absorb all available volume \cite{during2009breakdown}.  
\begin{definition}
    Entropy $S\equiv \int_{-\Delta}^{\Delta} \ln\left[ s(\omega) \right] d\omega =  2 \Delta  \left(\ln \left(\frac{\Theta }{\Delta ^2+\Xi ^2}\right)+2\right)-4 \Xi  \tan ^{-1}\left(\frac{\Delta }{\Xi }\right)$. 
\end{definition}
Hence and from Definition 4: $\frac{\partial S}{\partial \mu}=0$ \cite{wu2019thermodynamic}.

\begin{figure}
    \centering
    \includegraphics[scale=0.4]{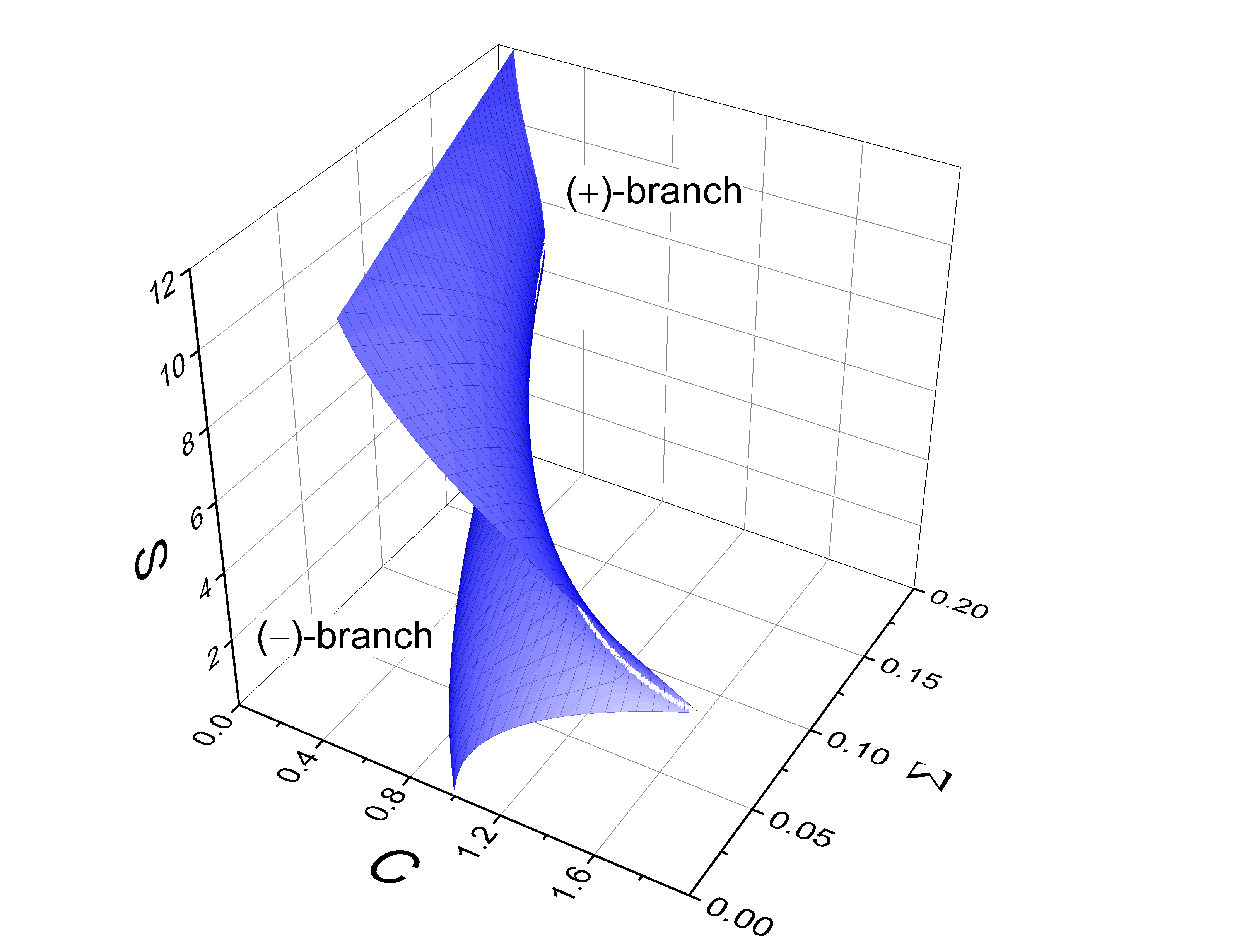}
    \caption{Dimensionless entropy $S$ for both DS-branches, $\Theta=60 \pi$ and $\chi=0$.}
    \label{fig:fig14}
\end{figure}

The dimensionless entropy is shown for both DS branches in Fig. \ref{fig:fig14}. The figure demonstrates lesser entropy for the ($-$)-branch. One may comment on that fact in the following way. The $P_0$-solution for a ($-$)-branch has a finite limit for $\zeta \to 0$ \cite{podivilov2005heavily} that is it is ``connectable'' with a classical soliton of the nonlinear Schr\"{o}dinger equation in the sense of \cite{haus1992analytic}. In other words, this branch is in a ``ground state'' and has no excited internal degrees of freedom, so its entropy is minimal. 

The ($+$)-branch is energy-scalable, i.e., it belongs to a DSR range. It has excitable internal degrees of freedom so that its entropy grows with an approach to the vacuum instability threshold, where it is maximal and grows with a temperature along the extreme DSR level $C=2/3$, $\Sigma=0$ as:

\begin{equation}\label{eq:maxS}
    S_{max}=\sqrt{\frac{7}{2}} \left(\ln \left(\frac{12 \Theta }{13}\right)+2\right)-\sqrt{\frac{10}{3}} \tan ^{-1}\left(\sqrt{\frac{21}{5}}\right),
\end{equation}
\noindent so that $\frac{\partial S_{max}}{\partial \Theta}=\frac{\sqrt{7/2}}{\Theta} \neq 0$. In particular, this ``high-entropy'' branch with enriched internal degrees of freedom has a larger ``informational capacity'' that could make DS a prospective tool for information transmission \cite{wan2023ultra}.  

Other thermodynamic values could be defined as \cite{wu2019thermodynamic}:
\begin{definition}
    Enthalpy (internal energy) $U \equiv \int_{-\Delta }^{\Delta } \frac{\Theta \, \omega ^2}{\Xi ^2+\omega ^2} \, d\omega,$
\end{definition}
\noindent so that $\frac{\partial S}{\partial U}=\frac{2}{\Theta}$.

\begin{definition}
    Energy (``particle number'' or energy contained in condensate) $\mathcal{E} \equiv \int_{-\Delta }^{\Delta } \frac{\Theta}{\Xi ^2+\omega ^2} \, d\omega,$
\end{definition}
\noindent so that $\frac{\partial S}{\partial \mathcal{E}}=\frac{-4 \mu}{\Theta} $.

\begin{definition}
    Gibbs free energy $\mathcal{F} \equiv U - \Theta \,S= -2 \Theta  \left(\Delta  \ln \left(\frac{\Theta }{\Delta ^2+\Xi ^2}\right)-\Xi  \tan ^{-1}\left(\frac{\Delta }{\Xi }\right)+\Delta \right),$
\end{definition}

\noindent so that the minimal free energy along the extreme DSR level $C=2/3$, $\Sigma=0$ is:

\begin{equation}\label{eq:minF}
    \mathcal{F}_{min}=\Theta\left(\sqrt{\frac{5}{6}}  \tan ^{-1}\left(\sqrt{\frac{21}{5}}\right)-\sqrt{\frac{7}{2}} \left(\log \left(\frac{12 \Theta }{13}\right)+1\right)\right).
\end{equation}

\begin{figure}
    \centering
    \includegraphics[scale=0.4]{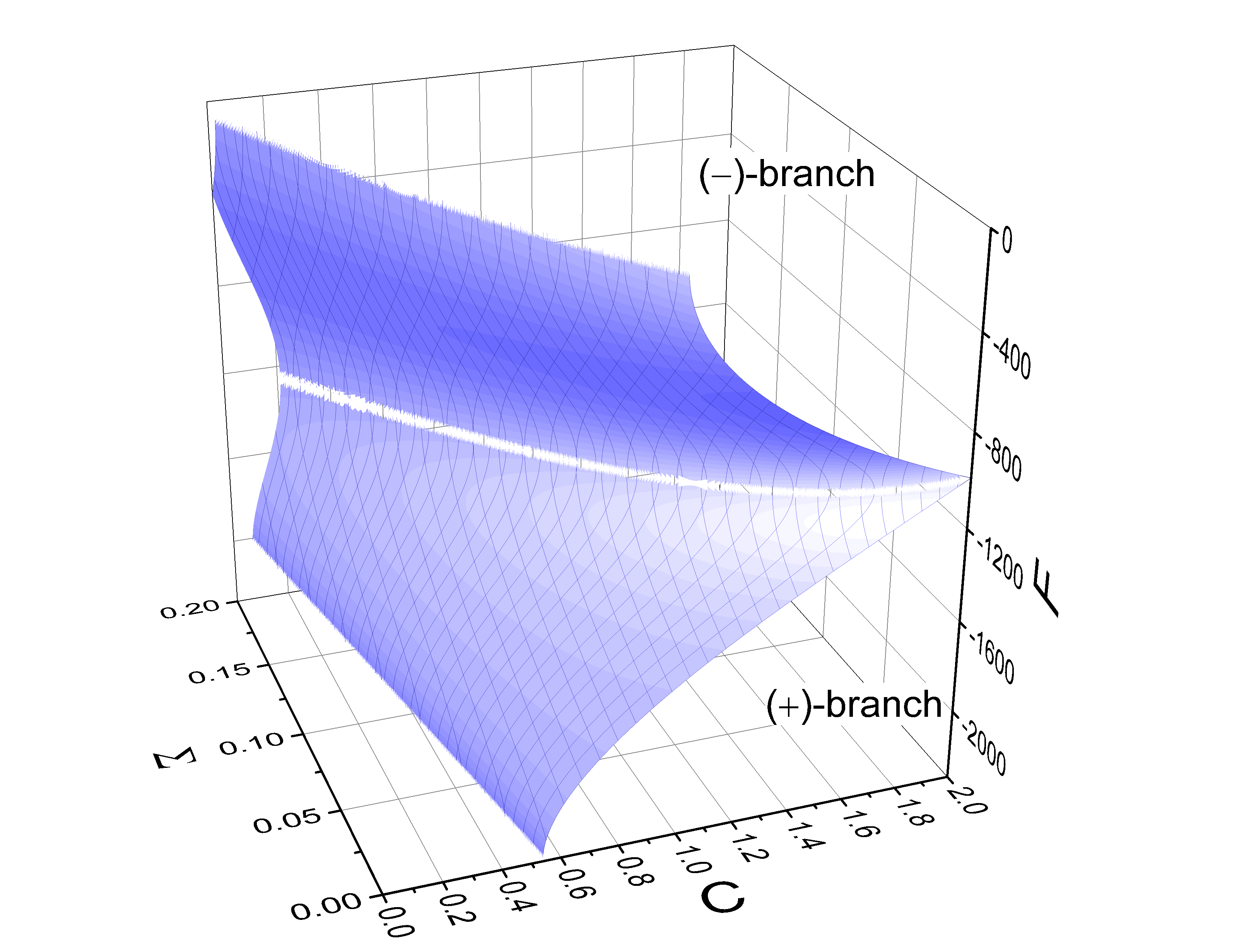}
    \caption{Dimensionless free energy $\mathcal{F}$ for both DS-branches, $\Theta=60 \pi$ and $\chi=0$.}
    \label{fig:fig15}
\end{figure}

The free energy is plotted in Fig. \ref{fig:fig15} for both DS branches. It is negative, i.e., DS is a thermodynamic preferable state within a range confined by the master diagram. It could be considered as an equilibrium state forming spontaneously from an incoherent basin. The energy-scalable (DSR) branch has the lowest free energy in the vicinity of the vacuum instability border and decreases with $\Theta$ (\ref{eq:minF}). Such minimization of free energy agrees with the analogous feature of BEC. 

\section{Conclusion}\label{sec12}

We have presented the adiabatic theory of a dissipative soliton (DS). It is based on the assumption that DS is strongly chirped, which requires domination of the nondissipative factors, such as  Kerr nonlinearity and GDD, over the dissipative ones, such as self-amplitude modulation and spectral dissipation. The complex cubic-quintic nonlinear Ginzburg-Landau equation (CQGLE) could describe all these factors. Under spatio-temporal duality, CQGLE can represent a broad range of nonlinear dynamical phenomena, particularly optical DS and weakly dissipative BEC. As CQGLE is not integrable in the general form, the approximated approaches to its study are highly desirable. The adiabatic approximation restrains a range of CQGLE parameters but keeps them remarkably realistic. Meanwhile, the obtained solutions are general within this range in the sense that they do not fix the relations between the equation parameters. This class of solutions belongs to the single-parametric family \cite{kharenko2011highly} that associates them with ``true'' solitons.

One of the advantages is that the obtained solutions are formulated in a spectral domain that allows for tracing the close analogies with the kinetic theory approaches to an interpretation of DS characteristics. The analytical expressions are straightforward in the case of vanishing imaginary quintic term. The DS spectrum has the shape of a truncated Lorentzian function so that all spectra can be divided into flat-top and finger-like classes. The division between them is defined by the equality of the truncation frequency and the Lorentzian width. These values play the role of two correlation lengths representing the internal DS phase inhomogeneity so that their equality is a markup of the maximal DS fidelity terms of its compressibility and the transition to the energy-scalable regime.

The latter corresponds to the DSR region, where DS is asymptotically scalable. The model provides simple analytical expressions corresponding to the DSR conditions. Advantageously, the concept of DSR is embedded organically into a representation of the DS parametric space in the form of two- (or three for a nonzero imaginary quintic term) dimensional master diagram, which connects a dimensionless DS energy and a parameter relating spectral and nonlinear dissipation to GDD and phase nonlinearity. The confined region of the last parameter corresponding to DSR has a simple analytical expression. The master diagram has a physically sound structure, which includes the stability threshold against vacuum instability, the region of DSR, a curve of maximum fidelity, etc. Moreover, the signatures characterizing a transition to the DSR regime are explicitly visible in the experiment as a transition to a constant spectral width, a Lorentzian peak in the spectrum, and a change of the DS squeezing to its broadening. All these phenomena are closely analogous to BEC.

The Rayleigh-Jeans spectral shape of DS and two independent correlation scales that diverge with the DS energy scaling suggest that a strongly chirped DS is akin to an incoherent (or partially coherent) soliton. The latter can be treated as an ensemble of interacting ``quasi-particles'' confined by a collective potential \cite{picozzi1}. Indeed, the analysis demonstrates that DS has a nontrivial internal structure so that such ``particles'' or their conglomerates can be excited, which perturbs the DS spectral and temporal profiles but preserves its total integrity. In some cases, this leads to DS turbulence.

The internal kinetics of DS allows the application of a thermodynamic language so that DS can be characterized by temperature, chemical potential, entropy, and free energy. The adiabatic theory expresses these values through the DS and CQNGLE parameters and demonstrates the thermodynamic differences between two types of DS ``populating'' the master diagram. Also, the thermodynamic viewpoint connects a limit of the DS energy scalability with the vanishing of chemical potential and the internal entropy growth.

We demonstrate that two branches of the DS solutions, namely, satisfying and non-satisfying DSR conditions, have different thermodynamical properties. Both types of DS have negative free energy that manifests an enhanced tendency for their formation from a basin. Also, the free energy under the DSR condition is minimal, while the entropy is maximal. The last can be interpreted as a manifestation of the nontrivial internal structure of such a type of DS. One may suppose that excitation of this structure could limit the DS energy scalability, that is, break DSR.

We believe that the approaches presented in this work will be helpful in the different areas, including photonics and BEC. In particular, an explicit definition of the DS energy-scalability limit can be expressed thermodynamically. The closely connected and unexplored problem is the analysis of the DS-basin interaction, which is essential to understanding the DS self-emerging \cite{gat2005light}. Also, including the higher-order derivative terms in CQGLE describing, in particular, higher-order GDD, is of interest from the viewpoint of the DS chaotization and the distortion of its internal structure.  
\backmatter

\section*{Declarations}

\bmhead{Ethics approval and consent to participate} Not applicable.
\bmhead{Consent for publication} All authors (VLK, AR, ES, ITS) consent to publication of this Work.
\bmhead{Availability of data and materials} Data underlying the results presented in this paper are not publicly available at this time but may be obtained from the authors upon reasonable request.
\bmhead{Competing interests} The authors (VLK, AR, ES, ITS) declare no conflicts of interest.
\bmhead{Funding} The work is supported by the Norwegian Research Council projects \#303347 (UNLOCK), \#326503 (MIR), and by ATLA Lasers AS.
\bmhead{Authors' contributions} The authors (VLK, AR, ES, ITS) contributed equally to this Work. 
\bmhead{Acknowledgements} The work of VLK, AR and ITS was supported by NFR projects \#303347 (UNLOCK), \#326503 (MIR), and by ATLA Lasers AS (ES).

\begin{appendices}

\section{Factorization}\label{secA1}

For $\chi=0$, the equation for a spectral deviation is:

\begin{equation}\label{eq:A1}
    \frac{d}{d t} \Omega\! \left(t\right)=\frac{\left(\frac{\kappa \beta \left(\Omega\left(t\right)^{2}-\Delta^{2}\right) \left(-\gamma-\zeta \beta \left(\Omega\left(t\right)^{2}-\Delta^{2}\right)\right)}{\gamma^{2}}-\sigma-\alpha \Omega\! \left(t\right)^{2}\right) \left(\Omega\! \left(t\right)^{2}-\Delta^{2}\right)}{\beta \left(3 \Omega\! \left(t\right)^{2}-\Delta^{2}\right)}.
\end{equation}

\noindent Let us write the numerator of (\ref{eq:A1}) as

\begin{equation}\label{eq:A2}
    \frac{\frac{\kappa \beta \left(\Omega\left(t\right)^{2}-\Delta^{2}\right) \left(-\gamma-\zeta \beta \left(\Omega\left(t\right)^{2}-\Delta^{2}\right)\right)}{\gamma^{2}}-\sigma-\alpha \Omega\! \left(t\right)^{2}}{\beta}=\left(3 \Omega\! \left(t\right)^{2}-\Delta^{2}\right) \epsilon \left(\Xi^{2}+\Omega\! \left(t\right)^{2}\right)
\end{equation}

Collecting the coefficient before the powers of $\Omega(t)$ gives

\begin{gather}\label{eq:A3}
   \left(-\frac{\kappa \beta \zeta}{\gamma^{2}}-3 \epsilon\right) \Omega\! \left(t\right)^{4}+\left(\frac{\frac{\kappa \,\beta^{2} \Delta^{2} \zeta+\kappa \beta \left(-\gamma+\zeta \beta \,\Delta^{2}\right)}{\gamma^{2}}-\alpha}{\beta}+\Delta^{2} \epsilon-3 \epsilon \Xi^{2}\right) \Omega\! \left(t\right)^{2}-\nonumber \\
   -\frac{\frac{\kappa \beta \,\Delta^{2} \left(-\gamma+\zeta \beta \,\Delta^{2}\right)}{\gamma^{2}}+\sigma}{\beta}+\Delta^{2} \epsilon \Xi^{2}=0.
\end{gather}
\noindent Equating the coefficient to zero and taking into account $\Delta^2=\gamma P_0/\beta$ results in:
\begin{align}
    b=-\frac{1}{3} \frac{\kappa \beta \zeta}{\gamma^{2}},\\
    \beta \Xi^{2}=-\frac{5}{3} P_{0} \gamma+\frac{\gamma}{\zeta}+\frac{\gamma^{2} \alpha}{\kappa \beta \zeta},\\
    P_{0}=\frac{3}{4} \frac{1-\frac{1}{2} c\pm\sqrt{\left(1-\frac{1}{2} c\right)^{2}-\frac{4 \zeta \sigma}{\kappa}}}{\zeta}.
\end{align}

This decomposition allows excluding a singularity from the denominator in Eq. (\ref{eq:A1}) and, thereby, avoiding the nonphysical solutions. Such a procedure would also be highly desirable for (\ref{eq:tchirp}).

\section{Numerical calculation of the DS parameters}\label{secA2}
That is Matlab code for calculating the DS parameters of the complex cubic-quintic nonlinear Ginzburg-Landau equation.

\begin{lstlisting}
%%%%%%%%%%%%%%%%%%%%%%%%%%%%%%%%%%%%%
%   Cubic-quintic CGLE              %                  %
% x is the c parameter              %
% y is the soliton energy           %
% z is the spectral half-width      %
% Vladimir Kalashnikov              %
% kalashnikov@ntnu.no               %
%%%%%%%%%%%%%%%%%%%%%%%%%%%%%%%%%%%%%
clear
Nt = 1000;
chi = 5;% control parameter chi
Sigma = 0;% control parameter Sigma

for k=2:1000 
    c = 2 - 2*(k-1)/1000;
    
% Spectral half-width for the positive branch
eq8 = sqrt((0.2e1*(c/chi+3+4/chi)* ...
(c/0.2e1+0.3e1/0.2e1*chi+ ...
    0.2e1 + sqrt(((c - 2)^2 - 16*Sigma*...
    (c/chi + 1))))/(c/chi + 1) -...
    (3*c) - (9*chi) - (32/chi*Sigma) - 0.12e2)/...
    (c/chi + 1)*c)/0.4e1;
    
% Spectral half-width for the negative branch
eq9 = sqrt((0.2e1*(c/chi + 3 + 4/chi)*...
(c/0.2e1 + 0.3e1/0.2e1*chi +...
    0.2e1 - sqrt(((c - 2)^2 - ...
    16*Sigma*(c/chi + 1))))/(c/chi + 1) -...
        (3*c) - (9*chi) - (32/chi*Sigma) - ...
        0.12e2)/(c/chi + 1)*c)/0.4e1;
    
% Curve, where the branches merge
eq10 = sqrt((0.2e1*(c/chi + 3 + 4/chi)*...
(c/0.2e1 + 0.3e1/0.2e1*chi +...
    0.2e1 - 0*sqrt(((c - 2)^2 - 16*Sigma*...
    (c/chi + 1))))/(c/chi + 1) -...
        (3*c) - (9*chi) - (32/b*Sigma) - 0.12e2)/...
        (c/chi + 1)*c)/0.4e1;

Delta = eq8;

if (imag(Delta)==0)

z(k)=Delta;

domega = 2*Delta/Nt;

omega = [-Delta:domega:Delta]; 

 

ar = (sqrt(((c*chi + 4*Delta^2 -...
4*omega.^2)/c/chi)) - 0.1e1).* ...
    ((sqrt(((c*chi + 4*Delta^2 - ...
    4*omega.^2)/c/chi)) - 0.1e1)*c*...
        chi + (8*omega.^2) - (4*Delta^2))/c.*...
        (((c*chi + 4*Delta^2 - ...
        4*omega.^2)/c/chi).^(-0.1e1/0.2e1))./...
        ((Sigma*c + c*omega.^2+... 
    c*chi + c*chi^2 + chi*Delta^2 -...
    chi*omega.^2).*(sqrt(((c*chi + 4*... 
    Delta^2 - 4*omega.^2)/c/chi)) - 0.1e1) -...
    (2*(Delta^2 - omega.^2)*...  
    (chi + 1)))/0.2e1;
    
for kk=1:Nt
    if (ar(kk)<0)
            ar(kk)=0;  
    else
            ar(kk)=ar(kk);   
    end
end

arg = ar(2:Nt);

E = trapz(arg)*domega;

x(k) = c;
    if (imag(E)==0)    
    y(k) = E;
    else
        y(k) = 0;
    end
else
end
end
\end{lstlisting}

\end{appendices}


\bibliography{sn-bibliography}

\end{document}